\documentclass[seceq]{ptptex}

\usepackage{graphicx}

\title{
Relativistic Chiral Mean Field Model for Finite Nuclei%
}

\author{
Yoko \textsc{Ogawa,}\footnote{E-mail: ogaway@rcnp.osaka-u.ac.jp}
Hiroshi \textsc{Toki,}\footnote{E-mail: toki@rcnp.osaka-u.ac.jp}
Setsuo \textsc{Tamenaga}\footnote{E-mail: stame@rcnp.osaka-u.ac.jp}
and Akihiro \textsc{Haga}\footnote{E-mail: haga@rcnp.osaka-u.ac.jp}
}

\inst{
Research Center for Nuclear Physics (RCNP), Osaka University, \\
Ibaraki 567-0047, Japan
}

\recdate{April 21, 2009}

\abst{
We present a relativistic chiral mean field (RCMF) model, which is a method for the proper treatment of 
pion-exchange interaction in the nuclear many-body problem.  There the dominant term of the pionic correlation is
expressed in two-particle two-hole (2p-2h) states with particle-holes having pionic quantum number, $J^{\pi}$.  
The charge-and-parity-projected relativistic mean field (CPPRMF) model developed so far treats surface properties 
of pionic correlation in 2p-2h states with $J^{\pi}$ = 0$^{-}$ (spherical ansatz).  
We extend the CPPRMF model by taking 2p-2h states with higher spin quantum numbers, 
$J^{\pi} = 1^{+}, 2^{-}, 3^{+},$ ... to describe the full strength of the pionic correlation in the 
intermediate range ($r >$ 0.5 fm).  We apply the RCMF model to the $^{4}$He nucleus as a pilot calculation
for the study of medium and heavy nuclei.  We study the behavior of energy convergence with the 
pionic quantum number, $J^{\pi}$, and find convergence around $J^{\pi}_{\rm max} = 6^{-}$.  
We include further the effect of the short-range repulsion in terms of the unitary correlation operator 
method (UCOM) for the central part of the pion-exchange interaction.  The energy contribution of about 
50$\%$ of the net two-body interaction comes from the tensor part and 20$\%$ comes from 
the spin-spin central part of the pion-exchange interaction.
}

\begin{document}

\maketitle

\section{Introduction}
Chiral symmetry is the most important symmetry as a key to understanding the mechanism of mass generation 
and pion dynamics in the subatomic world governed by the strong interaction \cite{lee}.
At the hadron level, chiral symmetry is described using the linear $\sigma$ model introduced by Gell-Mann 
and Levy \cite{gellmann}.  The Nambu-Jona-Lasinio (NJL) model is used for the mass generation and chiral properties 
of quark matter \cite{nambu}.  
Pion emerges with a zero mass field as a Nambu-Goldstone boson by spontaneous chiral symmetry breaking.  
With a small explicit symmetry breaking term, the pion acquires a small finite mass as compared with other mesons.  
Pion plays an important role as a mediator of intermediate- and long-range parts of nucleon-nucleon interaction \cite{yukawa}.  
It is extremely interesting and challenging to construct a nuclear system with the chiral Lagrangian, where the pion plays 
an important role in the nuclear structure.  We therefore study the property of a nuclear system with the linear $\sigma$ model 
of Gell-Mann and Levy \cite{gellmann}.  The proper treatment of the pion-exchange interaction in a way consistent with 
the many-body description becomes a key issue.  For this purpose, it is necessary to construct a relativistic framework
describing naturally the mass generation of hadrons and at the same time treat the pion-exchange interaction.  
We would like to develop a theoretical framework to handle the pion-exchange interaction together with the partial 
restoration of chiral symmetry in the nuclear system. 

We have started this program of treating the chiral sigma model Lagrangian in terms of the relativistic mean field 
approximation by introducing parity-mixed intrinsic single-particle states \cite{toki1,ogawa1}.  We further perform
the projection of the parity and, in addition, the charge of the total wave function from 
the parity- and charge-symmetry-broken intrinsic nuclear wave function, which is named 
charge-and-parity-projected relativistic mean field (CPPRMF) model \cite{ogawa2,ogawa3}.   
We have shown that the pion plays an important role in generating an effect of spin-orbit splitting and leads to 
$jj$-closed shell magic numbers \cite{ogawa2,ogawa3}.  This charge and parity projection was also performed 
in the nonrelativistic Hartree-Fock scheme \cite{sugimoto, ikeda,myo1,myo2}.

The CPPRMF model provides a specific feature that the 2-particle states of 2p-2h component have a spatially 
compact structure due to the pseudo-scalar nature of the pion exchange interaction.  The 0p-0h to 2p-2h 
matrix elements of the pion-exchange interaction to provide this compact structure, and the energy variation 
with respect to the spatial wave function in the CPPRMF is extremely important.  This is a critical difference 
of the CPPRMF model from any perturbative treatment of pion-exchange interaction \cite{weise,arima,brown}.  
In the CPPRMF model, the dominant contribution of the pion is expressed in terms of 2p-2h configurations 
with particle-hole spin-parity to be $J^{\pi}=0^-$ (spherical ansatz). The pionic energy contribution per nucleon, 
however, decreases rapidly with mass number in the systematic calculation of various medium and heavy 
nuclei\cite{ogawa3}.  This result seems inconsistent with that of the chiral perturbation theory for nuclear matter
by Kaiser et al.\cite{weise}. They show that an iterated one pion-exchange of Hartree term gives a large 
energy contribution to the binding energy of nuclear matter.  Our energy systematics of the CPPRMF model also
seems to be not accordance with a result of the variational Monte-Carlo (VMC) calculation of light nuclei obtained 
by the Argonne-Illinois group\cite{pandharipande}.  This fact shows that the pion-exchange interaction has 
a volume effect.  Hence, it is an important next step to include the volume effect in the CPPRMF model.

In nuclear matter, the momentum is the good quantum number and, particularly, the pseudo-scalar nature 
of the pion exchange interaction produces the contribution of large momentum component.  
If we were to take into account the volume effect of the pion exchange interaction, we have to take 
angular momenta of the pion, $L_{{\cal P}}$, up to the value satisfying  $q_{\rm max} R=L_{{\cal P}}^{\rm max}$, 
where $R$ represents the nuclear radius and $q_{\rm max}$ represents the maximum momentum of an exchanged 
pion in pion-exchange interaction. If we take the maximum momentum as twice of 
$k_F\sim$1.4 fm$^{-1}$ \cite{brown,akaishi}, we ought to take the angular momentum of the pion up to 
$L_{{\cal P}}^{\rm max}\sim 6$ for $^4$He.  Hence, we have to extend the CPPRMF model to include higher 
angular momentum components of the pion. The angular momentum of the pion, $L_{{\cal P}}$, and the 
total spin-parity quantum number of the particle-hole states, $J^{\pi}$, are related as 
$0^{-} \otimes L_{{\cal P}}^{(-1)^{L_{{\cal P}}}} = J^{\pi}$, where $0^{-}$ is an intrinsic spin-parity of pions.  
Owing to this relation between both quantities, we call  the total spin-parity of the particle-hole states, 
$J^{\pi}$, ''pionic quantum number" hereafter. The pseudo-scalar nature of the pion-exchange interaction 
is expressed using the 2p-2h components with pionic quantum number, $J^\pi=0^-$, in the CPPRMF model so far.  
As stated above, it is natural to extend the CPPRMF model to include 2p-2h components with higher
pionic quantum number; $J^\pi=0^{-}, 1^{+}, 2^{-}, 3^{+},...$.  

In this paper, we formulate the relativistic chiral mean field (RCMF) model to treat full strength the pionic correlations 
in the nuclear many-body system.  The RCMF model is based on the mean field picture, where all the nucleons move 
in the mean field generated by other nucleons. The pion has the pseudo-scalar nature, which is then treated 
in 2p-2h components with pionic quantum number; $J^{\pi}=0^-, 1^+, 2^-, 3^+,...$.   Although the probability 
of nucleon pairs being brought up to high momentum states is of the order of $10\sim15 \%$, the energy gain 
associated with the pion exchange interaction is essential to bind nucleons in the nucleus.

This article is arranged as follows. In $\S$2, we discuss the formulation of the RCMF model for finite nuclei.  
In $\S$3, we apply the RCMF model to the $^{4}$He nucleus as a pilot calculation.  The numerical results 
are presented in this section and detailed discussions are made for the properties of the resulting wave functions 
and energy components.  A summary and  the outlook are presented in $\S$4.

\section{Formulation}
We develop here the RCMF model  for the description of finite nuclei.  The RCMF model is a natural extension 
of the CPPRMF model, which properly takes into account the pseudo-scalar nature of the pion-exchange
interaction. As discussed in the introduction, we include all possible 2p-2h configurations to take into 
account the full strength of the pionic correlations. We take the pion-exchange interaction with the 
form factor due to the finite size effect of the nucleon. We calculate explicitly the direct part of the 
pion-exchange interaction and drop exchange terms as the natural extension of the CPPRMF 
model \cite{ogawa2,ogawa3}.

\subsection{Chiral model Lagrangian}
We start with the linear sigma model Lagrangian of Gell-Mann and Levy, where the pion field appears symmetrically 
with the $\sigma$ field.  The pseudo-scalar pion-nucleon coupling in the linear $\sigma$ model leads to an unrealistically
large attractive contribution through the strong coupling between positive and negative energy states, and we have to 
treat properly the effect of the negative energy states.  We thus employ the nonlinear realization of chiral Lagrangian,  
which is obtained by the Weinberg transformation of the linear $\sigma$ model \cite{weinberg}.  We take the 
lowest order term in the pion field, and the Lagrangian density in the nonlinear representation is written 
as \cite{boguta, ogawa1}

\begin{equation}
{\cal L} = {\cal L}_{\sigma, \omega} + {\cal L}_{\pi},
\end{equation}
where
\begin{eqnarray}
{\cal L}_{\sigma, \omega} & = & \bar{\psi}
( i\gamma _{\mu}\partial^{\mu} - M 
- g_{\sigma}\sigma 
- g_{\omega}\gamma _{\mu}\omega^{\mu} ){\psi} \\ \nonumber
& + &   
\frac{1}{2}\partial _{\mu}\sigma\partial^{\mu}\sigma- \frac{1}{2}{m_{\sigma}}^2\sigma^2
- \lambda{f_{\pi}}\sigma^3 - \frac{\lambda}{4}\sigma^4 \\ \nonumber
& - &   
\frac{1}{4}\omega _{\mu\nu}\omega^{\mu\nu} + \frac{1}{2}{m_{\omega}}^2\omega _{\mu}\omega^{\mu} \\ \nonumber
& + & {\widetilde{g_{\omega}}}^2f_{\pi}\sigma \omega _{\mu}\omega^{\mu}
+ \frac{1}{2}{\widetilde{g_{\omega}}}^2\sigma^2 \omega _{\mu}\omega^{\mu},
\end{eqnarray}
and
\begin{equation}
{\cal L}_{\pi} =- \frac{g_{A}}{2f_{\pi}}
\bar{\psi} \gamma _{5}\gamma _{\mu}\partial^{\mu}\pi^{a}\tau^{a}
{\psi} + \frac{1}{2}\partial _{\mu}\pi^{a}\partial^{\mu}\pi^{a} - \frac{1}{2} {m_\pi}^2{\pi^a}^2.
\end{equation}
The nucleon and $\omega$ meson masses are obtained by $\sigma$ meson condensation in the vacuum.
The effective nucleon mass and the effective $\omega$ meson mass are given by 
$M^{\ast} = M +  g_{\sigma}\sigma$, and 
$m^{\ast}_{\omega} = m_{\omega} +  \widetilde{g_{\omega}}\sigma$.
Masses and coupling constants are set as $M = g_{\sigma} f_{\pi}$,
$m^{2}_{\pi} = \mu^{2} + \lambda f^{2}_{\pi}$, $m^{2}_{\sigma} =  \mu^{2} + 3\lambda f^{2}_{\pi}$, 
and $m_{\omega} = \widetilde{g_{\omega}}f_{\pi}$. We take the empirical values for masses and 
the pion decay constant as $M = 939$ MeV, $m_{\omega} = 783$ MeV, $m_{\pi} = 139$ MeV, 
and $f_{\pi} = 93$ MeV. The $\sigma$-nucleon coupling constant, $g_{\sigma}$, and the
$\sigma$-$\omega$ coupling constant, $\widetilde{g_{\omega}}$, are fixed by the relations, 
$g_{\sigma} = M / f_{\pi} = 10.1$ and $\widetilde{g_{\omega}} = m_{\omega} / f_{\pi} = 8.42$. 
The strengths of the $\sigma$ meson self-energy terms depend on the $\sigma$ meson mass, 
$m_{\sigma}$, through the relation $\lambda = (m^{2}_{\sigma} - m^{2}_{\pi}) / 2f^{2}_{\pi}$. 
The $\sigma$ meson mass and $\omega$-nucleon coupling constant, $g_{\omega}$, are the free
parameters. We introduce the pion-nucleon axial vector coupling constant, $g_{A}$, in this Lagrangian.  
We set $g_{A} = $ 1.25, which is related with the coupling strength in the free-space $\pi NN$ scattering 
by the Goldberger-Treiman relation \cite{goldberger}.

\subsection{Ground state wave function}
The total wave function of the whole nuclear system is written as
\begin{equation}
\Psi  = \Psi_{N} \otimes \Psi_{M}.
\end{equation}
The meson wave function, $\Psi_{M}$, is a coherent state for the $\sigma$ and $\omega$ mesons.
As for the nucleon wave function, we write the projected wave function of CPPRMF model,
\begin{equation}
\Psi=N_0 | 0p-0h \rangle + \sum_{i}N_i | 2p-2h  \rangle_i +
\sum_{k} N_k | 4p-4h  \rangle_k+ \cdot\cdot\cdot \hspace{0.5cm}.
\end{equation}
The contribution of the pionic correlation is dominantly expressed in terms of 2p-2h states, where particle-holes 
of 2p-2h states have $J^\pi=0^-$.  For $^4$He, the probability of 4p-4h states is negligibly small \cite{ogawa2}. 
Hence, we neglect the configurations with more than 4p-4h states.  We  introduce higher multipoles by coupling 
particle-holes of 2p-2h states to various spins and parities with pion quantum number.  It is a straightforward 
extension to introduce higher mutipoles for the pionic quantum number, $J^\pi$.  Hence, we would like to 
write the total wave function for the nuclear part, $\Psi_N$, as 0p-0h state and the combination of 2p-2h states,
\begin{equation}
|\Psi _{N} \rangle  = \alpha _{0} | 0p-0h ; b_{0} \rangle + \sum_{i} \alpha _{i} | 2p_{i}-2h_{i} ; b_{i} \rangle,
\end{equation}
where $i$ denotes various 2p-2h states.  The normalization condition of the total wave function is 
$\langle \Psi_{N} | \Psi_{N} \rangle = \alpha_{0}\alpha^{\ast}_{0} + \sum_{i}\alpha_{i}\alpha^{\ast}_{i} = 1$.
The amplitude, $\alpha _{i}$, represents the weights of 2p-2h states. They are variational parameters for 
total energy minimization. The 0p-0h wave function, $| 0p-0h ; b_{0} \rangle$, represents the relativistic mean 
field (RMF) ground state, which is obtained by solving the RMF equations for Hamiltonian, $\hat{H}_{\sigma, \omega}$.  
Nucleons occupy mean field levels up to the Fermi surface. The 2p-2h wave function is constructed as
\begin{equation}
|  2p_{i}-2h_{i}  ; b_{i} \rangle = 
\Bigl{[} 
| { p-h ; JMTM_{T}(b_i) } \rangle \otimes | { p'-h' ; J^{'}M^{'}T^{'}M'_T(b_i)}\rangle 
\Bigr{]}^{(00)}_i.
\end{equation} 
 
\begin{figure}[t]
\centerline{\includegraphics[width=6.0cm,clip]{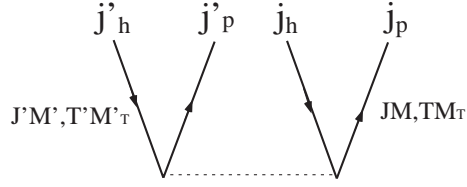}}
\caption{\label{fig1} 2p-2h Hartree diagram.}
\end{figure}

A single-particle wave function for 2p-2h states is written as
\begin{eqnarray}
\psi_{njm\tau} &=&
\left(\begin{array}{@{\,}r@{\,}}
   iG_{n\kappa}(r){\cal Y}_{\kappa m}(\hat{r},\sigma) \\
    F_{n\kappa}(r){\cal Y}_{\bar{\kappa} m}(\hat{r},\sigma)  \\
\end{array}\right)\zeta(\tau),
\end{eqnarray}
where $G_{n \kappa}(r)$ and $F_{n\kappa}(r)$ represent the radial parts of the upper 
and lower components of the Dirac spinor. The orbital angular momentum and spin state are given as 
${\cal Y}_{\kappa m}(\hat{r},\sigma)
= \sum_{m_{l}, m_{s} }( l m_{l} 1/2 m_{s} | j m)Y_{l m_{l}}(\hat r) \chi_{m_{s}}(\sigma)$.
The label $n$ distinguishes radial wave functions of states with the same $\kappa, m$ quantum number. 
The label $\kappa$ defines $l$ and $j$ as $\kappa=-(l+1)$ for $l=j-1/2$ and $\kappa=l$ for 
$l=j+1/2$, and $\bar{\kappa} = -\kappa$ represents the opposite parity state.  $\zeta(\tau)$ 
denotes isospin state. We follow conventions as in Ref.\citen{bjorken}. 

The upper radial part is given by Gaussian function as,
\begin{equation}
G_{n\kappa} 
=  {\cal N}N(l, b_{i}) r^{l} {\rm exp} \Bigl{(} -\frac{r^{2}}{2b^{2}_{i}} \Bigr{)}, \hspace{0.2cm}
N(l, b_i)= \sqrt{ \frac{2}{b^{2l+3}_{i}\Gamma(\frac{2l+3}{2})}},
\end{equation}
where $b_{i}$ denotes a range of Gaussian function and ${\cal N}$ is the normalization factor of the spinor, 
$\psi_{njm\tau}$. In the case that the particle state has the same quantum number as the hole state, 
we take the next form to make the particle state orthogonal to the hole state for the upper radial part,
\begin{equation}
G_{n\kappa} 
= ({\cal N}_{1} + {\cal N}_{2} r^{2}) N(l, b_{i}) r^{l} {\rm exp} \Bigl{(}- \frac{r^{2}}{2b^{2}_{i}} \Bigr{)},
\end{equation}
where ${\cal N}_{1}$ and ${\cal N}_{2}$ are fixed by the normalization and orthogonalization 
with the hole state.  As for the lower radial part, we take the following form, 
\begin{equation}
F_{n\kappa} = \frac{1}{2M} \Big{(} \frac{d}{dr} + \frac{\kappa+1}{r} \Big{)} G_{n\kappa},
\end{equation}
which is the relation of the lower radial component with the upper radial component 
for plain wave function for small momentum, $E\sim M$. We take this assumption to minimize 
the number of variational parameters.

\subsection{Hamiltonian}
In the relativistic mean field approximation, the source term of the pion disappears for parity-conserved 
single-particle states, and therefore, the solution obtained at the mean field level does not include the effect 
of the pion.  We take this relativistic mean field ground state as an initial state for the RMF 0p-0h ground state.  
We express 2p-2h states in terms of Gaussian functions with various range parameters for pionic correlations.

The relation between the Hamiltonian density and the Lagrangian density is written as
\begin{equation}
{\cal H} = \sum_{\phi} \frac{\partial{\cal L}}{\partial \dot{\phi}} \dot{\phi} - {\cal L},
\end{equation}
where $\phi$ denotes the nucleon field, $\psi$, and $\pi$, $\sigma$, $\omega$ meson fields. 
The Hamiltonian density for $\sigma$ and $\omega$ fields is
\begin{eqnarray}
{\cal H} _{\sigma, \omega} &=& \bar{\psi} ( -i \vec{\gamma} \cdot \vec{\nabla} + M 
+ g_{\sigma}\sigma
+ g_{\omega} \gamma_{\mu}\omega^{\mu} )\psi \\ \nonumber
&+& \frac{1}{2}\vec{\nabla}\sigma \cdot \vec{\nabla}\sigma + \frac{1}{2}m^{2}_{\sigma}\sigma^{2}
+\lambda f_{\pi} \sigma^{3} + \frac{\lambda}{4} \sigma^{4} \\ \nonumber
&- & \frac{1}{2}\vec{\nabla}\omega^{0} \cdot \vec{\nabla}\omega^{0} 
+ \frac{1}{2}( \vec{\nabla} \times \vec{\omega} ) \cdot (\vec{\nabla} \times \vec{\omega} )
- \frac{1}{2} m^{2}_{\omega}\omega_{\mu} \omega^{\mu} \\ \nonumber
&-&\widetilde{g}^{2}_{\omega} f_{\pi} \sigma \omega_{\mu} \omega^{\mu} 
- \frac{1}{2} \widetilde{g}^{2}_{\omega} \sigma^{2} \omega_{\mu} \omega^{\mu},
\end{eqnarray}
where we consider the static case, and the time derivative terms, $\partial_{0}\phi$, vanish.
We take the mean field approximation for $\sigma$ and $\omega$ field operators,
\begin{eqnarray}
\sigma &\longrightarrow& \langle \Psi | \sigma | \Psi \rangle \equiv \bar{\sigma}, \\
\omega &\longrightarrow& \langle \Psi | \omega _{\mu} | \Psi \rangle \equiv \delta_{\mu, 0} \bar{\omega}_{0}.
\end{eqnarray}
The expectation value of the spatial component, $\langle \vec{\omega} \rangle $, vanishes under the rotational 
symmetry condition. Hereafter, we write just '$\sigma$' and '$\omega$' as the classical fields, $\bar{\sigma}$ 
and $\bar{\omega} _{0}$, respectively.  The Hamiltonian for $\sigma$ and $\omega$ mesons is obtained as  
 \begin{eqnarray}
 \hat{H}_{\sigma, \omega} &=& \int d^{3}x {\cal H}_{\sigma, \omega} \\ \nonumber
 &=& \int d^{3}x 
 {\Big\{} \psi^{\dagger}
 (-i\vec{\alpha} \cdot \vec{\nabla} + \gamma_{0} (M+ g_{\sigma}\sigma) + g_{\omega}\omega )
 \psi + {\cal E}_{\rm meson}(\sigma, \omega) {\Big\}},
 \end{eqnarray}
 where
 \begin{eqnarray}
 {\cal E}_{\rm meson}(\sigma, \omega) 
 &=& \frac{1}{2}(-\vec{\nabla}^{2} + m^{2}_{\sigma})\sigma^{2} + \lambda f_{\pi}\sigma^{3} 
 + \frac{1}{4}\lambda \sigma^{4} \\ \nonumber
 & -& \frac{1}{2}(-\vec{\nabla}^{2} + m^{2}_{\omega}) \omega^{2}
 - \widetilde{g}^{2}_{\omega}f_{\pi}\sigma \omega^{2} -\frac{1}{2}\widetilde{g}^{2}_{\omega} \sigma^{2} \omega^{2}.
 \end{eqnarray}
The Hamiltonian density for the pion part is given by
\begin{equation}
{\cal H}_{\pi} =\frac{g_{A}}{2f_{\pi}} \bar{\psi}  \gamma _{5}\vec{\gamma}\cdot \vec{\nabla}\pi^{a} \tau^{a} \psi
+ \frac{1}{2}\vec{\nabla}\pi^{a} \cdot \vec{\nabla}\pi^{a} + \frac{1}{2} {m_{\pi}}^{2}\pi^{a}\pi^{a}.
\end{equation}
Using the Eular-Lagrange equation, 
\begin{equation}
\partial _{\mu}\frac{\partial {\cal L}}{\partial(\partial_{\mu} \pi^{a})} - \frac{\partial {\cal L}}{\partial \pi^{a}} = 0,
\end{equation}
we obtain the Klein-Gordon equation for the pion field as
\begin{equation}
(-\vec{\nabla}^{2} + {m_{\pi}}^{2}) \pi^{a} = \frac{g_{A}}{2f_{\pi}} \vec{\nabla} 
\cdot \bar{\psi}\gamma _{5}\vec{\gamma} \tau^{a} \psi ,
\end{equation}
where $a = 0, \pm$.
From this Klein-Gordon equation, we obtain the integral form for the pion field,
\begin{eqnarray}
\pi^{a}(\vec{x}) 
&=&
\frac{g_{A}}{2f_{\pi}} \int d^{3} y \frac{e^{-m_{\pi}|\vec{x} - \vec{y}|}}{4\pi |\vec{x} -\vec{y}|} \vec{\nabla}_{y} \cdot
\bar{\psi}(\vec{y})\gamma _{5} \vec{\gamma}\tau^{a}\psi(\vec{y})  \\ \nonumber
&=&
-\frac{g_{A}}{2f_{\pi}} \int d^{3}y 
\bar{\psi}(\vec{y})\gamma _{5} \vec{\gamma}\tau^{a}\psi(\vec{y}) \cdot \vec{\nabla}_{y}
\int \frac{d^{3}q}{(2\pi)^{3}} \frac{e^{i \vec{q} \cdot (\vec{x} -\vec{y})}}{{\vec{q}}^{2} + {m_{\pi}}^{2}}.
\end{eqnarray}
By using this integral form and the Klein-Gordon equation of the pion field, the Hamiltonian for the pion-exchange interaction
is written as
\begin{eqnarray}
\hat{H}_{\pi} &=& \int d^{3}x {\cal H}_{\pi} \\ \nonumber
&=&
\frac{1}{2} 
\int \frac{d^{3}q}{(2\pi)^{3}}
\int \int d^{3}x d^{3}y 
\bar{\psi}(\vec{x})\gamma _{5}\vec{\gamma} \cdot \vec{q} \tau^{a}\psi(\vec{x}) \\ \nonumber
&&\hspace{2.5cm}\times
\Big{[} - {\big(} \frac{g_{A}}{2f_{\pi}} {\big)}^{2}
\frac{e^{i \vec{q} \cdot (\vec{x} -\vec{y})}}{{\vec{q}}^{2} + {m_{\pi}}^{2}}
\Big{]}
\bar{\psi}(\vec{y})\gamma _{5}\vec{\gamma}\cdot \vec{q} \tau^{a}\psi(\vec{y}).
\end{eqnarray}
The total Hamiltonian is given by
\begin{equation}
\hat{H} = \hat{H}_{\sigma, \omega} + \hat{H}_{\pi}.
\end{equation}

\subsection{Pionic energy minimization}
We take $\Psi$ as the trial wave function with variational parameters, $\alpha_{i}$, $b_{i}$, $\sigma$, 
and $\omega$. As for the nucleon part, we search for the energy minimum within the ($\alpha _{i}$, $b_{i}$)-plane.
For this purpose, we solve the following two-energy minimization equations simultaneously.  One is given by
\begin{equation}
\frac{\partial}{\partial \alpha _{i}} \langle \Psi | \hat{H} - E | \Psi \rangle = 0, 
\end{equation}
and the other is given by
\begin{equation}
\frac{\partial}{\partial b _{i}} \langle \Psi | \hat{H} - E | \Psi \rangle = 0.
\end{equation}
The first energy minimization condition provides the variational parameter $\alpha _{i}$.  The second one 
is very important from the viewpoint of the pseudo-scalar nature of the pion-exchange interaction. 
The finding obtained by CPPRMF method is that the particle states of 2p-2h states have compact distributions 
due to the high-momentum components caused by the pseudo-scalar nature. In the second step of minimization, 
we can achieve the full strength of pionic correlations by using a small number of bases.  We make "pionic optimization"
in this size variation.

We obtain equations for the meson mean fields by varying the total energy with respect to $\sigma$ and $\omega$ 
classical fields as
\begin{equation}
\frac{\partial}{\partial \phi }\langle \Psi | \hat{H} | \Psi \rangle = 0,
\end{equation}
where $\phi$ means $\sigma$ and $\omega$. Differential equations for the $\sigma$ and $\omega$ meson fields are
\begin{eqnarray}
(-\vec{\nabla}^{2} + m^{2}_{\sigma})\sigma 
&=& - g_{\sigma}\rho_{s}
- 3\lambda f_{\pi} \sigma^{2} - \lambda \sigma^{3}
+\widetilde{g_{\omega}}^{2} f_{\pi}  \omega^{2}
+\widetilde{g_{\omega}}^{2} \sigma \omega^{2}, \\
(-\vec{\nabla}^{2} + m^{2}_{\omega} )\omega 
&=& g_{\omega} \rho_{v} 
-2\widetilde{g_{\omega}}^{2} f_{\pi} \sigma \omega
-\widetilde{g_{\omega}}^{2} \sigma^{2} \omega,
\end{eqnarray}
where the scalar and vector densities are calculated using the 0p-0h and 2p-2h wave functions presented in Eq. (2.6) as
\begin{eqnarray}
\rho_{s} &=& \langle \Psi_N | : \bar{\psi} \psi : | \Psi _N\rangle, \\ \nonumber
\rho_{v} &=& \langle \Psi_N | : \bar{\psi} \gamma_{0}  \psi : | \Psi_N \rangle.
\end{eqnarray}
We solve these four equations given by energy functional variations, Eqs. (2.24), (2.25), (2.27), (2.28), (2.29), 
self-consistently. We note here that these meson fields are affected by the 2p-2h pionic correction terms 
through each density shown in Eq. (2.29). The 0p-0h RMF ground state is re-decided by these $\sigma$ and 
$\omega$ meson fields for every iterative calculation until we obtain the self-consistent solution.

\subsection{Matrix element of the pion-exchange interaction}
We show how to calculate the matrix element, $\langle 2p_{i}-2h_{i} ; b_{i} | \hat{H}_{\pi} |  0p-0h ; b_{0} \rangle$,
in this subsection. At first, we consider particle-hole states induced by the pionic response with quantum number, 
$JMTM_{T}$, and two particle-hole states couple to 0$^{+}$ ground state.  We put the suffix, $"p"$, for the 
quantum numbers of the particle state like $n_{p}, j_{p}, l_{p}$, and we put the suffix, $"h"$, for the quantum numbers
of the hole state like $n_{h}, j_{h}, l_{h}$.  $l_{p}$ and  $l_{h}$ represent the single-particle orbital angular momentum
of the upper component of particle and hole states, respectively.  The lower components are represented by 
$\bar{l}_{p}$ and $\bar{l}_{h}$, respectively. The pion carries momentum $\vec{q}$ between two nucleons.  
Following the method shown in Ref. \citen{toki2}, we write down the matrix elements of the particle-hole excitation
induced by the pionic response as follows, 
\begin{eqnarray}
&&\langle p-h ; JMTM_{T} | 
\gamma_{0}\gamma_{5}\vec{\gamma} \cdot \vec{q} \tau_{\lambda} {\rm exp}(-i\vec{q} \cdot \vec{r}) 
| \hat{0} \rangle \\ \nonumber
&&\hspace{1cm}= iq \delta_{T1} \delta_{M_{T}\lambda} 
\Bigl{\{}
\langle (l_{p} \frac{1}{2})j_{p}(l_{h} \frac{1}{2})j_{h} ; JM | 
\vec{\sigma} \cdot \hat{q} {\rm exp}(-i \vec{q} \cdot \vec{r}) | \hat{0} \rangle \\ \nonumber
&&\hspace{3cm}
+ 
\langle (\bar{l_{p}} \frac{1}{2})j_{p}(\bar{l_{h}} \frac{1}{2})j_{h} ; JM | 
\vec{\sigma} \cdot \hat{q} {\rm exp}(-i \vec{q} \cdot \vec{r}) | \hat{0} \rangle
\Bigr{\}} \\ \nonumber
&&\hspace{1cm}= iq\delta_{T,1} \delta_{M_{T},\lambda} {\rm Y}^{\ast}_{JM}(\hat{q})
\Bigl{\{} 
\sum_{ L} a_{LJ} F^{JL\ast}_{\rm ph}(q) + \sum_{\bar{L}} a_{J \bar{L}} \bar{F}^{J \bar{L}\ast}_{\rm ph}(q) \Bigr{\}},
\end{eqnarray}
where for the upper component,
\begin{eqnarray}
F^{JL}_{\rm ph}(q) &=& 2 \langle LS ; J | j_{p} j_{h} ; J \rangle (-i)^{L} \sqrt{4\pi(2l_{h} + 1)}(l_{h} 0  L 0 | l_{p} 0 ) \\ \nonumber
&\times&\int^{\infty}_{0} r^{2} dr j_{L}(qr) G_{n_{p} j_{p} l_{p}}(r) G_{n_{h} j_{h} l_{h}} (r), \\ \nonumber
 \langle  LS ; J | j_{p} j_{h} ; J \rangle 
 &=& \Bigl{[} (2L+1)(2S+1)(2j_{p} +1)(2j_{h} +1) \Bigr{]}^{1/2} \\ \nonumber
&\times& \left\{
  \begin{array}{ccc}
              l_{p} & l_{h} & L \\
              \frac{1}{2}   & \frac{1}{2}   & S \\
              j_{p} & j_{h} & J  \\
  \end{array}
\right\}, \\ \nonumber
a_{JL} &\equiv& (-1)(J0 10 | L0) =
 \left\{
 \begin{array}{r}
              \Big( \frac{J}{2J+1} \Big)^{1/2}  \hspace{5mm} {\rm for} \hspace{5mm}  L = J-1, \\
              -\Big( \frac{J+1}{2J+1} \Big)^{1/2}  \hspace{5mm} {\rm for} \hspace{5mm} L = J+1. \\
  \end{array}
  \right.
\end{eqnarray}
and for the lower component, 
\begin{eqnarray}
\bar{F}^{J\bar{L}}_{\rm ph}(q) &=& 2 \langle \bar{L}\bar{S} ; J | j_{p} j_{h} ; J \rangle (-i)^{\bar{L}} \sqrt{4\pi(2\bar{l_{h}} + 1)}
(\bar{l_{h}} 0 ; \bar{L} 0 | \bar{l_{p}} 0 ) \\ \nonumber
&\times&\int^{\infty}_{0} r^{2} dr j_{\bar{L}}(qr) F_{n_{p} j_{p} \bar{l_{p}}}(r) F_{n_{h} j_{h} \bar{l_{h}}} (r), \\ \nonumber
\langle  \bar{L}\bar{S} ; J | j_{p} j_{h} ; J \rangle 
&=& \Bigl{[} (2\bar{L}+1)(2\bar{S}+1)(2j_{p} +1)(2j_{h} +1) \Bigr{]}^{1/2} \\ \nonumber
&\times&\left\{
\begin{array}{ccc}
              \bar{l_{p}} & \bar{l_{h}} & \bar{L} \\
              \frac{1}{2}   & \frac{1}{2}   & \bar{S} \\
              j_{p} & j_{h} & J  \\
\end{array}
\right\}, \\ \nonumber
a_{J\bar{L}} &\equiv& (-1)(J0 10 | \bar{L}0) =
\left\{
\begin{array}{r}
              \Big( \frac{J}{2J+1} \Big)^{1/2} \hspace{5mm} {\rm for} \hspace{5mm} \bar{L} = J-1, \\
    		  -\Big( \frac{J+1}{2J+1} \Big)^{1/2} \hspace{5mm} {\rm for} \hspace{5mm} \bar{L} = J+1. \\
\end{array}
\right.
\end{eqnarray}
$|\hat{0} \rangle$ represents the 0p-0h RMF ground state. Using this pionlike response matrix element, 
we construct the matrix element of the direct term for the correlation between 0p-0h and 2p-2h states as follows,
\begin{eqnarray}
&&\frac{1}{2}
\int \frac{d^3 q}{(2\pi)^3}
\langle {2p-2h} | \gamma_{0}\gamma_{5}\vec{\gamma} \cdot \vec{q} \tau_{\lambda}
\Big{[} - (\frac{g_{A}}{2f_{\pi}})^{2} 
\frac{{\rm exp}(i\vec{q} \cdot (\vec{r}_{1} - \vec{r}_{2}))}{m^{2}_{\pi} + \vec{q}^{2}}\Big{]}
\gamma_{0}\gamma_{5}\vec{\gamma} \cdot \vec{q}\tau_{\lambda}| \hat{0} \rangle \\ \nonumber
&&\hspace{2cm}= \frac{1}{2}\Big{(}- (\frac{g_{A}}{2f_{\pi} })^{2} \Big{)}
\sum_{MM'}(JMJ'M' |00) \sum_{M_{T}M'_{T}} (TM_{T}T'M'_{T}|00) \sum_{\lambda} (-1)^{\lambda} \\ \nonumber
&&\hspace{2cm}\times
\int \frac{d^3 q}{(2\pi)^{3}} \frac{1}{m^{2}_{\pi}+\vec{q}^{2}}
\langle p-h; JMTM_{T} | 
\gamma_{0}\gamma_{5}\vec{\gamma} \cdot \vec{q} \tau_{\lambda}{\rm exp}(i\vec{q} \cdot \vec{r}_{1}) 
| \hat{0} \rangle \\ \nonumber
&&\hspace{4.1cm}\times
\langle p-h ; J'M'T'M'_{T} | 
\gamma_{0}\gamma_{5}\vec{\gamma} \cdot \vec{q} \tau_{-\lambda}{\rm exp}(-i\vec{q} \cdot \vec{r}_{2}) 
| \hat{0} \rangle \\ \nonumber
&&\hspace{2cm}=
\frac{1}{2}(-1)^{J+1}\sqrt{3(2J+1)}\big{(} - (\frac{g_{A}}{2f_{\pi}})^{2} \big{)} \\ \nonumber
&&\hspace{2cm}\times
\int^{\infty}_{0}\frac{q^{2}dq}{(2\pi)^{3}}\frac{q^{2}}{m^{2}_{\pi}+\vec{q}^{2}} \Bigl{\{} 
\sum_{ L} a_{LJ} F^{JL\ast}_{\rm ph}(q) +
\sum_{\bar{L}} a_{J \bar{L}} \bar{F}^{J \bar{L}\ast}_{\rm ph}(q) \Bigr{\} } \\ \nonumber
&&\hspace{5.3cm}\times
\Bigl{\{ } 
\sum_{ L'} a_{L'J} F^{JL}_{\rm ph}(q) + \sum_{\bar{L'}} a_{J \bar{L'}} \bar{F}^{J \bar{L'}}_{\rm ph}(q)
\Bigr{\} }.
\end{eqnarray}
We shall include the pion form factor to take into account the finite nucleon size effect by replacing 
the pion propagator as \cite{machleidt}
\begin{equation}
\frac{1}{\vec{q}^{2} + m^{2}_{\pi}} 
\longrightarrow 
\frac{1}{\vec{q}^{2} + m^{2}_{\pi}} 
\Big{(}\frac{\Lambda^{2} - m^{2}_{\pi}}{\vec{q}^{2} + \Lambda^{2}} \Big{)}^{2},
\end{equation}
where $\Lambda$ represents the momentum cutoff parameter. 

\begin{figure}[b]
\centerline{\includegraphics[width=9.0cm,clip]{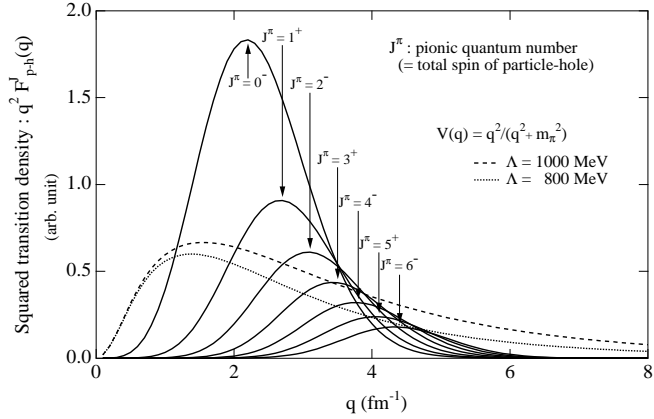}}
\caption{\label{fig2} Momentum-dependent part of pion exchange interaction, $\frac{q^2}{m_\pi^2+q^2}$, 
together with the form factor, and the squared particle-hole transition density, $q^{2} F^{J}_{\rm ph}(q)$, 
as functions of $q$. The squared particle-hole transition densities are shown for various $J^\pi$ by solid curves 
and the pion exchange interaction is shown by a dashed curve for $\Lambda=1000$ MeV and by a dotted curve 
for $\Lambda=800$ MeV. The parameters used for this presentation are $\sigma$ meson mass, 
$m_{\sigma}$ = 850 MeV and the $\omega$-nucleon coupling constant, $g_{\omega}$ = 7.07.}
\end{figure}
We would like to show here the integrand of the momentum integral of 0p-0h and 2p-2h matrix elements 
to see the connection of the pion quantum number in finite system and pion momentum in infinite matter.  
Shown in Fig. \ref{fig2} is the momentum-dependent part of the pion-exchange interaction, 
$\frac{q^2}{m_\pi^2+q^2}$, together with the form factor, and the squared particle-hole transition density, 
$q^{2} F^{J}_{\rm ph}(q)$, as functions of momentum, $q$.  Here, $q^{2}$ is the factor coming from 
the phase space, and we multiply this factor by $F^{J}_{\rm ph}(q)$ shown in Eq. (2.35).
\begin{eqnarray}
F^J_{\rm ph}(q) &=& 
\Bigl{\{} 
\sum_{ L} a_{LJ} F^{JL\ast}_{\rm ph}(q) + \sum_{\bar{L}} a_{J \bar{L}} \bar{F}^{J \bar{L}\ast}_{\rm ph}(q) 
\Bigr{\} } \\ \nonumber
&\times&
\Bigl{\{ } 
\sum_{ L'} a_{L'J} F^{JL}_{\rm ph}(q) + \sum_{\bar{L'}} a_{J \bar{L'}} \bar{F}^{J \bar{L'}}_{\rm ph}(q)
\Bigr{\} }.
\end{eqnarray}
The $0^-$ component has the largest transition density and peaks at smaller momenta, while the 
large $J^\pi$ components have gradually smaller transition densities and peak at larger momenta.  
To take into account the full strength of the pion-exchange interaction in the intermediate-range region, 
we have to include these higher spin pionic states.  The CPPRMF model takes into account the pionic 
strength indicated here by $J^\pi=0^-$. 

\section{Numerical results and discussions}
We apply the RCMF model to the $^{4}$He nucleus as a pilot calculation.  As the first step, we prepare 0p-0h RMF 
ground state, $(0s_{1/2})^{4}$, by solving the RMF equation obtained from the $\sigma$-$\omega$ Hamiltonian, 
$\hat{H}_{\sigma, \omega}$. We adjust the $\sigma$ meson mass, $m_{\sigma}$, and the $\omega$-nucleon 
coupling constant, $g_{\omega}$, so as to reproduce the root-mean-square (rms) matter radius of $^{4}$He, 
$R_{\rm rms} =$ 1.488 fm. We use two ranges for Gaussian function with b$_{0}$ = 0.7 and 1.4 fm, where the latter 
range is the dominant range.  We keep these two ranges during the self-consistent calculation.
For the $\sigma$ and $\omega$ meson fields, we expand them in the same way as the hole state using the two-range 
Gaussian functions,
\begin{equation}
\phi(r) = \sum_{\nu = 1, 2} C^{\phi}_{\nu} 
\sqrt{\frac{2^{5/2}}{b^{3}_{0, \nu}\Gamma(\frac{3}{2})}}
{\rm exp} \Big{(} -\frac{r^{2}}{b^{2}_{0, \nu}} \Big{)},
\end{equation}
where $\phi$ denotes $\sigma$ and $\omega$ meson fields.

\subsection{One-range Gaussian for particle states}
\begin{figure}[t]
\centering
\includegraphics[width=6.9cm,clip]{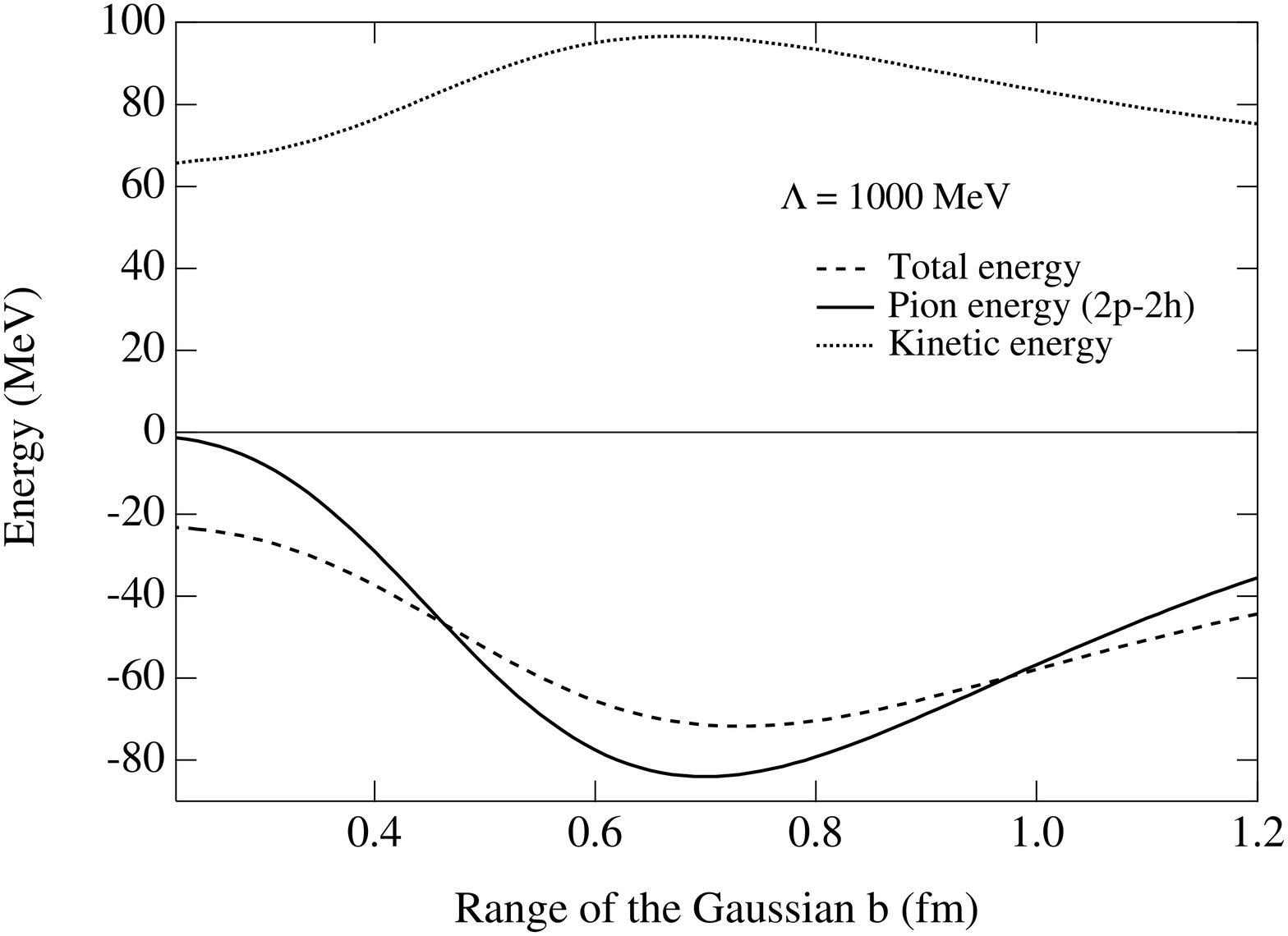}
\includegraphics[width=6.9cm,clip]{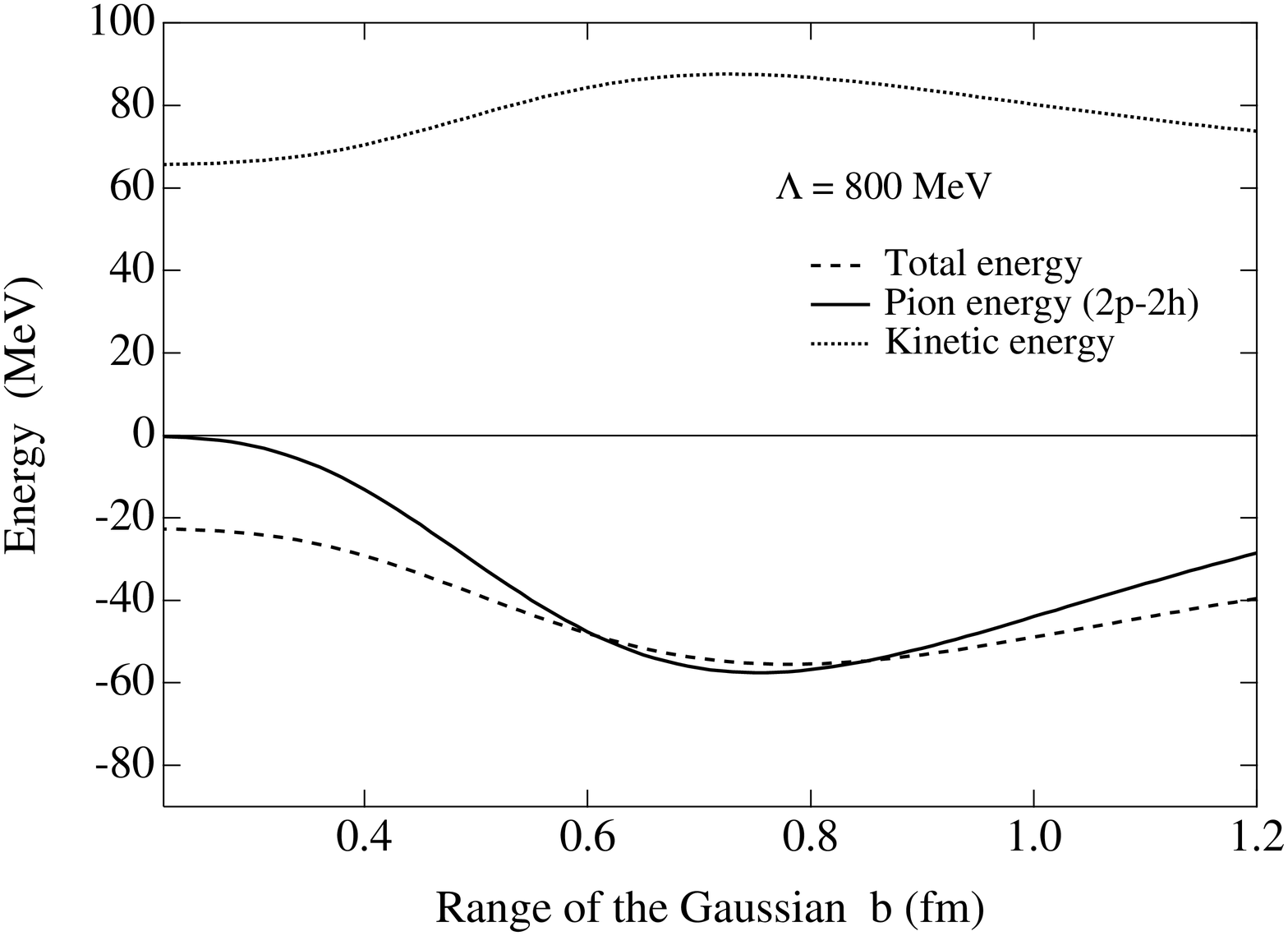}
\caption{\label{fig3} Total energy and various energy components as functions of the Gaussian range of 
particle state, $b$, for $^{4}$He with $J^{\pi}_{\rm max}=6^{-}$.  We take two cases for the cutoff 
momentum $\Lambda=1000$ MeV (left figure) and $800$ MeV (right figure).  The other free parameters, 
$\sigma$ meson mass and $\omega$-nucleon coupling constant, are 850 MeV and 7.07, respectively}
\end{figure}

In Fig. \ref{fig3}, we show the total energy of $^{4}$He and various energy components as a function of 
the Gaussian range of particle state, $b$. We take 2p-2h states up to $J^{\pi}_{\rm max} = 6^{-}$ of the 
pionic quantum number.  The total energy minimum is achieved at around $b= 0.8 \sim$ 0.7 fm, which corresponds 
to about half the Gaussian range, 1.4 fm, which is the dominant range of the hole state.  The particle state has a 
spatially compact distribution reflecting high-momentum components in the wave function due to the pseudo-scalar
nature of the pion-exchange interaction, $(\vec{\sigma}_{i} \cdot \vec{q})(\vec{\sigma}_{j} \cdot \vec{q})$. 
This important feature has been pointed out in CPPHF \cite{sugimoto} and CPPRMF \cite{ogawa2} methods 
in the case of the spherical pion field ansatz, $J^{\pi} = 0^{-}$. It is indispensable to have this wide variational space 
for 2p-2h states to take into account properly the pseudo-scalar nature of the pion-exchange interaction.  
Reflecting the pseudo-scalar nature, there is a strong correlation between kinetic energy and pion energy.   
The central energy, which is the summation of the scalar and vector potentials, is almost constant, which is not shown
in Fig. \ref{fig3}.  The energy contribution from the nonlinear term of $\sigma$ and $\omega$ mesons 
is also almost constant. We show the results of two cases for cutoff momentum of the pion form factor.  
The contribution of the pionic energy is large for large cutoff momentum, $\Lambda$.

\begin{figure}[t]
\centering
\includegraphics[width=6.9cm,clip]{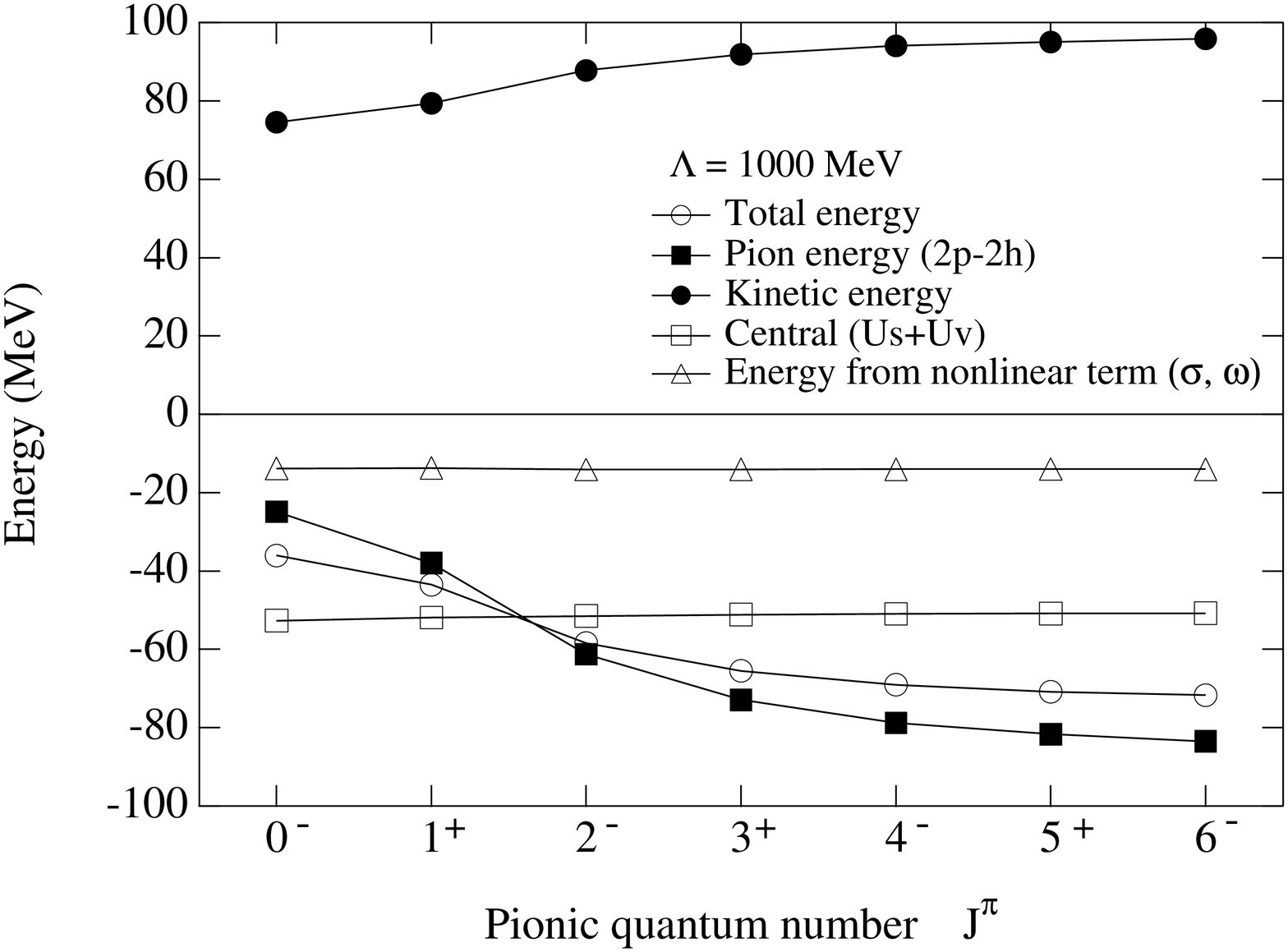}
\includegraphics[width=6.9cm,clip]{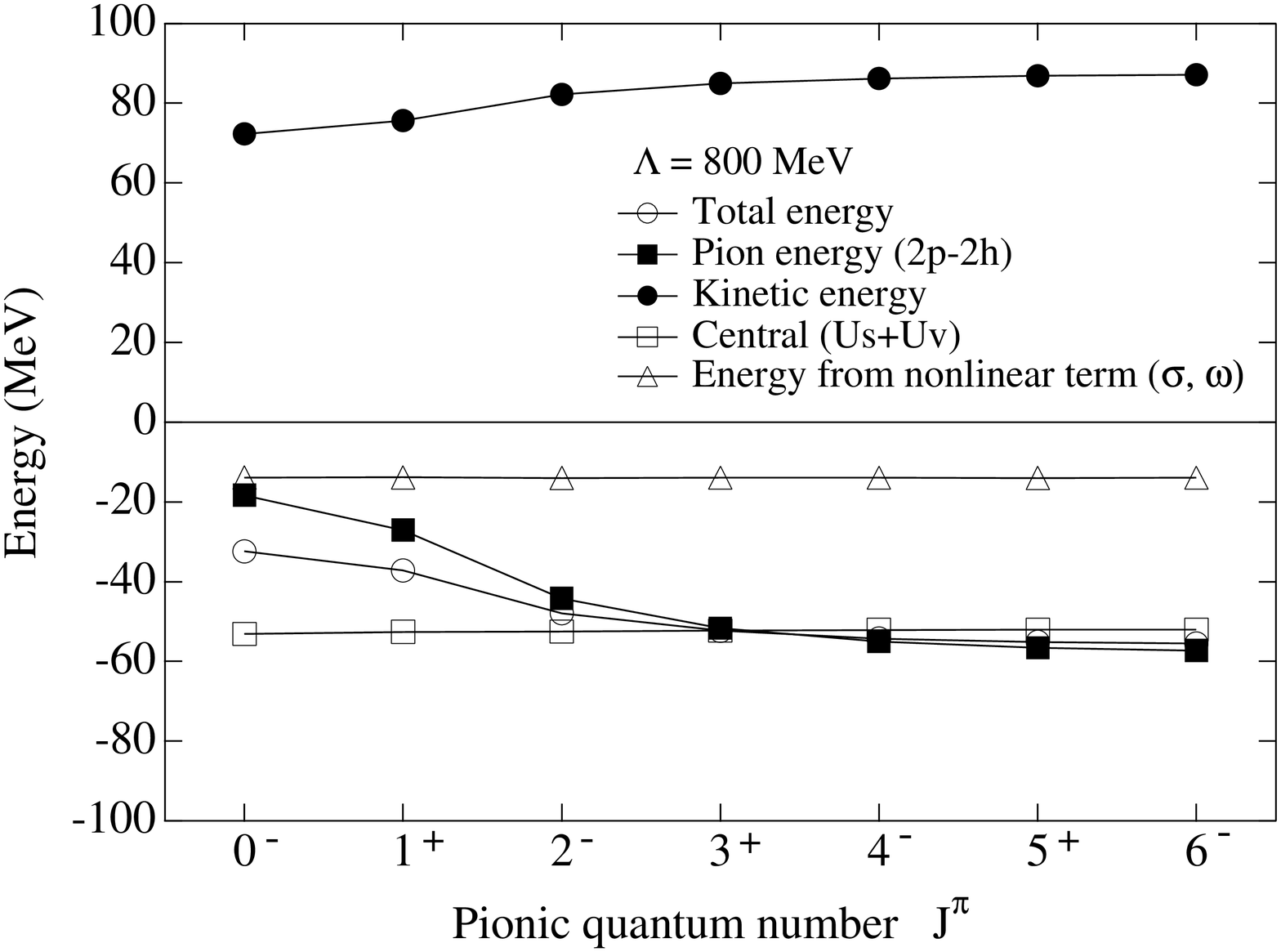}
\caption{\label{fig4} Total energy and various energy components as functions of pionic quantum number,  
$J^{\pi}$, for $^{4}$He. The cutoff momenta, $\Lambda$, are taken to be 1000 MeV (left figure) and 
800 MeV (right figure).  The other free parameters, $\sigma$ meson mass and $\omega$-nucleon coupling
constant, are 850 MeV and 7.07, respectively}
\end{figure}

In Fig. \ref{fig4}, we show the total energy and its components for $^{4}$He as functions of the pionic quantum number, 
$J^{\pi}$, for $\Lambda$ = 1000 MeV in the left-hand panel and $\Lambda$ =800 MeV in the right-hand panel.  
We see that the total energy and its components tend to converge as it approaches $J^{\pi} = 6^{-}$, which 
corresponds to the single-particle orbital angular momentum, $l_{\rm p}$ = 7.  These high-multipole components are 
required to take into account the full strength of the intermediate-range part of the relative distance of two-nucleon pairs 
due to the pion-exchange interaction for $^{4}$He.  The range parameter, which makes total energy minimum, changes 
from 0.83 fm with $J^{\pi}_{\rm max} = 0^{-}$ to 0.75 fm with $6^{-}$.   

\subsection{Multirange Gaussian expansion for particle states}
\begin{figure}[t]
\centerline{\includegraphics[width=8.0cm,clip]{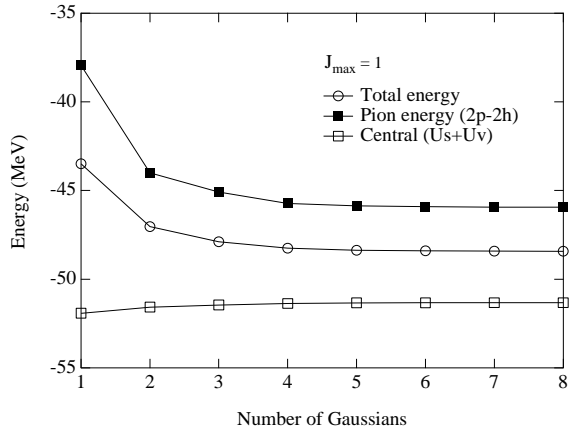}}
\caption{\label{fig5} Convergence of the total energy with the number of Gaussians for $^{4}$He.  
The pionic quantum number is taken up to $J^{\pi}_{\rm max}$ = 1$^{+}$. The cutoff momentum, $\Lambda$, 
is taken to be 1000 MeV. Two free parameters, $\sigma$ meson mass and $\omega$-nucleon coupling constant, 
are 850 MeV and 7.07, respectively.}
\end{figure}

We show in Fig. \ref{fig5} the total energy and the pion and central energies as functions of the number of 
Gaussian functions.  We search the energy minimum by changing the Gaussian range, $b_{i}$, for each step 
where we add a new Gaussian function.  We increase progressively a new Gaussian function until the energy 
convergence is realized.  We obtain energy convergence by using a relatively small number ($\sim 6$) of 
Gaussian function.  This is an advantage of the pionic optimized method.  Because the range parameter, $b_{i}$, 
is a variational parameter, we can minimize the number of range parameters in the above method.
Almost all range parameters, $b_{i}$, are found to be less than 1.0 fm.  In actual calculations, we take 8 ranges, 
whose values are $b_i =$ 0.4, 0.6, 0.8, 0.9, 1.0, 1.1, 1.4, 2.0 fm.  We include mostly small-size Gaussians 
and one large size, $b=2.0$, to take care of noncompact components.  The pion energy increases by around 20$\%$ 
compared with the case of the one-range Gaussian.  In this figure, we take up to $J^{\pi}$ = 1$^{+}$, where channels 
of particle states, $(p_{1/2})^{2}$, $(p_{3/2})^{2}$, and $(1s_{1/2})(d_{3/2})$, are taken into account, where 
the hole state is $(0s_{1/2})^2$.  Hereafter, we do not write explicitly the hole state.  

\subsection{Nuclear size and energy}
\begin{figure}[ht]
\centerline{\includegraphics[width=8.0cm,clip]{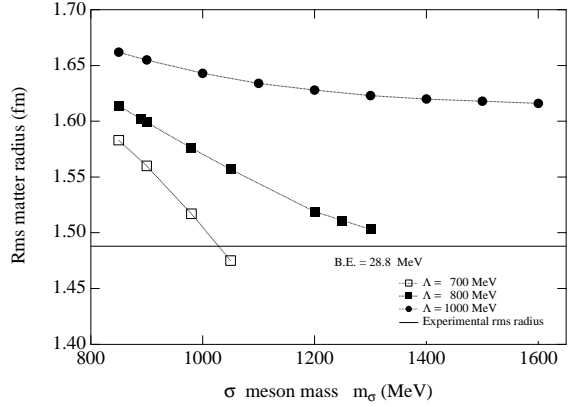}}
\caption{\label{fig6} Root-mean-square matter radius of $^4$He as a function of the $\sigma$ meson mass 
for three cases of the cutoff momentum, $\Lambda$.  The result of $\Lambda$ = 1000 MeV is shown 
by the solid circle, that of $\Lambda$ = 800 MeV by the solid square, and that of $\Lambda$ = 700 MeV 
by the open square.  The pionic quantum number is taken up to $J^{\pi}_{\rm max}$ = 6$^{-}$.  
The number of Gaussians is 8, where $b_i$ = 0.4, 0.6, 0.8, 0.9, 1.0, 1.1, 1.4, 2.0 fm.}
\end{figure}

We show in Fig. \ref{fig6} the rms matter radius of $^4$He as a function of the $\sigma$ meson mass 
in three cases with the cutoff momenta, $\Lambda$ = 700, 800, and 1000 MeV.  We adjust the $\omega$-nucleon 
coupling constant to reproduce the empirical binding energy of $^{4}$He without the Coulomb interaction.
As the cutoff momentum, $\Lambda$, becomes larger, a heavier $\sigma$ meson mass is required to obtain 
the rms matter radius.  When the $\sigma$ meson mass becomes larger, the attractive force decreases, while 
the pion-exchange interaction contributes more to the binding energy.  As a consequence, the equation of state 
becomes softer, as the cutoff momentum becomes larger.  In the case of $\Lambda$ = 1000 MeV, the rms matter radius 
gradually becomes flat as the $\sigma$ meson mass.  Even if we take a significantly large $\sigma$ mass, it is not possible
to find $m_{\sigma}$ to reproduce the binding energy and matter radius simultaneously.  In cases of $\Lambda$ = 700 
and 800 MeV, the $\sigma$ meson mass under the realization of the saturation property is significantly large compared 
with that (500 $\sim$ 600 MeV) of the Walecka model \cite{walecka,sugahara} and that (850 MeV) obtained under 
the spherical pion field ansatz, $J^{\pi} = 0^{-}$, in the CPPRMF method \cite{ogawa2}.  In the case where the matter 
radius and total energy are reproduced, we find that the bound state cannot be realized only by scalar and vector potentials.  
The bound state is realized by the attraction almost entirely due to the pion-exchange interaction.

\begin{figure}[t]
\centerline{\includegraphics[width=8.0cm,clip]{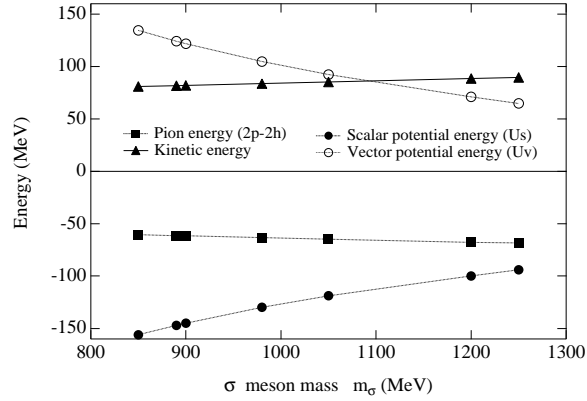}}
\caption{\label{fig7} Various energy components for $^{4}$He as functions of the $\sigma$ meson mass, 
m$_{\sigma}$, with the cutoff momentum of $\Lambda$ = 800 MeV.  The $\omega$-nucleon coupling constant, 
$g_{\omega}$, is adjusted to reproduce the empirical binding energy of $^{4}$He without the Coulomb interaction. 
The pionic quantum number is taken up to $J^{\pi}_{\rm max}$ = 6$^{-}$.  
The number of Gaussians is 8, where $b_i =$ 0.4, 0.6, 0.8, 0.9, 1.0, 1.1, 1.4, 2.0 fm.}
\end{figure}
In Fig.\ref{fig7}, we show various energy components of $^{4}$He as functions of the $\sigma$ meson mass.  
As the $\sigma$ meson mass becomes heavier, its field strength decreases and the attraction due to 
the $\sigma$ meson is reduced.  Instead, the energy contribution due to the pion-exchange interaction increases.
The reduction of the strength of the $\sigma$ meson is accompanied by a decrease in the strength of 
the $\omega$ meson field.  As a consequence, the strength of the spin-orbit force becomes smaller as 
the $\sigma$ meson mass increases. This fact indicates the importance of pions in the nucleus on the formation 
of the shell structure (spin-orbit splitting effect) \cite{myo1, arima, ogawa1}.  Around 70$\%$ of the total attractive 
potential comes from the pion-exchange interaction at the $\sigma$ meson mass, $m_{\sigma}$ = 1250 MeV 
in the case of $\Lambda=800$ MeV.  This amount of energy contribution due to the pion-exchange interaction 
is in good agreement with the result of the variational Monte-Carlo (VMC) method for light nuclei ($A <10$) obtained
by the Argonne-Illinois group \cite{pandharipande}.   They point out that the energy contribution of the pion-exchange
interaction to the net two-body attraction is 70$\sim$80$\%$. 

\subsection{Tensor and spin-spin central parts of pion-exchange interaction}
We new discuss the contributions of the tensor part and spin-spin central part of the pion-exchange interaction.    
The pion-exchange interaction is separated into the spin-spin part and tensor part as
\begin{equation}
\frac{\vec{\sigma}_{1} \cdot \vec{q} \vec{\sigma}_{2} \cdot \vec{q}
\vec{\tau}_{1} \cdot \vec{\tau}_{2}}{\vec{q}^{2} + m^{2}_{\pi}}
= \frac{1}{3} \Big{(}
\vec{\sigma}_{1} \cdot \vec{\sigma}_{2} \frac{\vec{q}^{2}}{\vec{q}^{2} + m^{2}_{\pi}}
+\frac{3\vec{\sigma}_{1} \cdot \vec{q} \vec{\sigma} _{2}\cdot \vec{q} 
- \vec{q}^{2}\vec{\sigma}_{1} \cdot \vec{\sigma}_{2} }{\vec{q}^{2} + m^{2}_{\pi}} \Big{)}
\vec{\tau}_{1} \cdot \vec{\tau}_{2} .
\end{equation}
The first term is the spin-spin central part and the second term is the tensor part.   

\begin{figure}[b]
\centering
\includegraphics[width=7.5cm,clip]{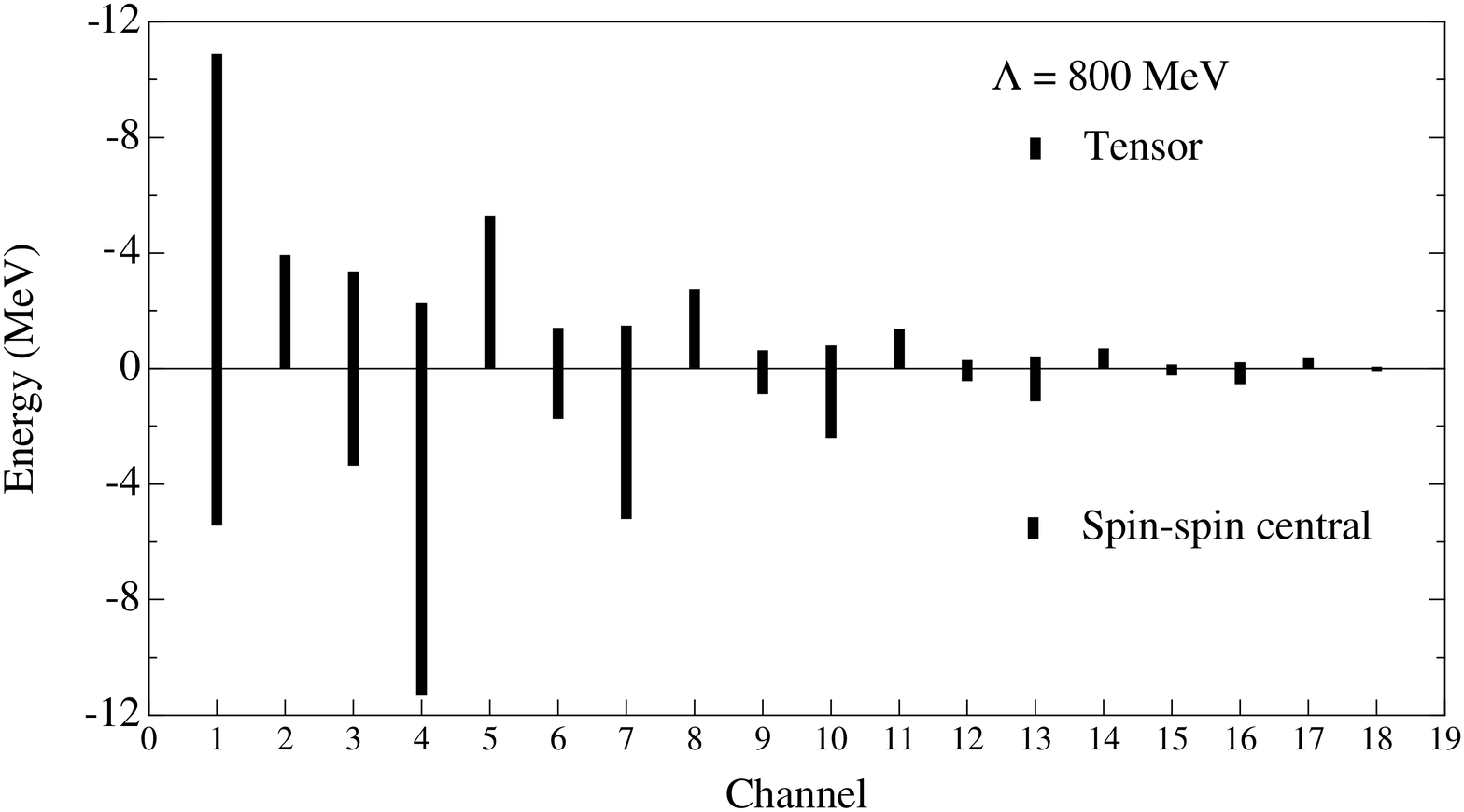}
\includegraphics[width=6.0cm,clip]{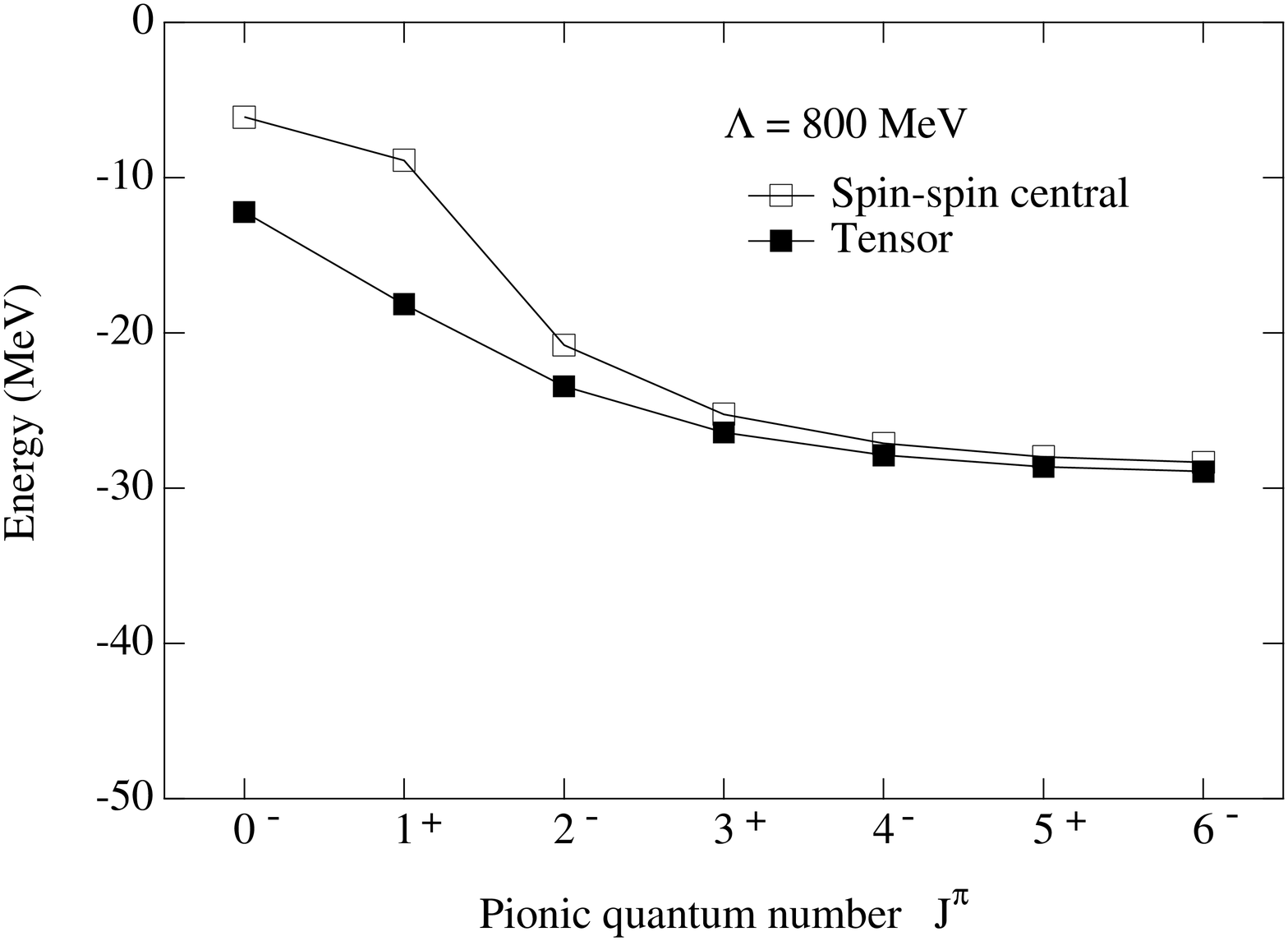}
\caption{\label{fig8} 
Energy contributions of the tensor part and spin-spin central part of the pion-exchange interaction 
for various 2p-2h states. In the left-hand panel shown in the upper half is the tensor part and in the lower half 
is the spin-spin central part obtained with $\Lambda$ = 800 MeV.  In the right-hand panel is shown the energy 
contributions of the tensor part and spin-spin part of the pion-exchange interaction as functions of $J^\pi$.  
The numbers shown in the horizontal axis of the left-hand panel represent 2p-2h states; 
1:$(p_{1/2})^{2}$, 2:$(1s_{1/2})(d_{3/2})$, 3:$(d_{3/2})^{2}$, 4:$(p_{3/2})^{2}$, 
5:$(p_{3/2})(f_{5/2})$, 6:$(f_{5/2})^{2}$, 7:$(d_{5/2})^{2}$, 8:$(d_{5/2})(g_{7/2})$, 
9:$(g_{7/2})^{2}$, 10:$(f_{7/2})^{2}$, 11:$(f_{7/2})(h_{9/2})$, 12:$(h_{9/2})^{2}$, 
13:$(g_{9/2})^{2}$, 14:$(g_{9/2})(i_{11/2})$, 15:$(i_{11/2})^{2}$, 
16:$(h_{11/2})^{2}$, 17:$(h_{11/2})(j_{13/2})$, 18:$(j_{13/2})^{2}$. 
The number of Gaussians is 8, where $b_i$ = 0.4, 0.6, 0.8, 0.9, 1.0, 1.1, 1.4, 2.0 fm.}
\end{figure}

We show the energy contributions from the tensor part (upper half of left-hand panel) and spin-spin 
part (lower half of left-hand panel) of the pion-exchange interaction for various 2p-2h states in Fig. \ref{fig8}.  
In the right-hand panel, we show the energy contributions of the tensor part and spin-spin central part 
of the pion-exchange interaction as functions of $J^\pi$.  The most important 2p-2h configuration
is the $(p_{1/2})^{2}$ state.  This channel corresponds to the case of the spherical pion field ansatz, 
$J^{\pi} = 0^{-}$, which is only the configuration taken into account in the CPPRMF model \cite{toki1,ogawa2,ogawa3} 
and also some other works \cite{sugimoto, ikeda, myo1} in the nonrelativistic version.  The tensor part behaves 
similarly to that determined by Myo et al. \cite{myo2}.  On the other hand, we see a large contribution of the 
$(p_{3/2})^{2}$ state due to the spin-spin central interaction.  This spin-spin central contribution being large 
is not seen in other calculations \cite{myo2}.
In our results, the tensor and spin-spin central parts give almost the same amount of energy contribution for 
$\Lambda$ = 800 MeV as shown in the right-hand panel of Fig. \ref{fig8}.   About 40$\%$ of the total attractive
potential comes from the tensor part, and about 30$\%$ comes from the spin-spin central part. We compare 
these energy contributions with the result obtained by the Argonne-Illinois group \cite{pandharipande}.  The energy 
contribution of the tensor part to the net two-body attractive force is approximately 50$\%$ and that of the spin-spin 
central part is just approximately 20$\%$.  From this comparison, we can conclude that the energy gain due to the 
tensor part is underestimated by around 10$\%$; on the other hand, that due to the spin-spin central part is 
overestimated by around 10$\%$ in our calculation.  

The reason why we have overestimated the energy due to the spin-spin central part is that we do not consider
at all the effect of the short-range repulsion, which was pointed out by Jastrow \cite{jastrow}, which cuts down 
the wave function in the region less than 0.5 fm in the relative distance of two nucleons.  The effect of the 
short-range repulsion is serious for the spin-spin central part, because the spin-spin central part works 
when the change in the relative orbital angular momentum between two nucleons is 0, $\Delta L$ = 0.  
Since there is no centrifugal barrier, its amplitude becomes large, as two nucleons approach each other.  
On the other hand, the tensor part works when the change in the relative orbital angular momentum is 2, 
$\Delta L$ = 2. The relative wave function of a two-nucleon pair for the tensor case is pushed outside 
by the centrifugal potential.  

We consider that the problem is related to the overestimation of the spin-spin central part.  We have not 
treated at all the effect of the short-range repulsion.  If we perform the correct procedure for taking the 
many-body correlation in the short-range part, we may avoid the overestimation of the spin-spin central part.  
As a suitable procedure, the unitary operator correlation method (UCOM) is well known \cite{feldmeier} 
to describe the short-range correlation in the framework based on the single-particle picture.  
Once we can suppress the contribution of the spin-spin central part, we can increase the contribution 
of the tensor part by taking a larger cutoff momentum, $\Lambda$.  We discuss UCOM in the following section.

\subsection{Inclusion of the effect of the short-range repulsion }
We follow the method of UCOM for the spin-spin central part of the pion-exchange interaction, 
which is the first term in Eq. (3.2) as
\begin{equation}
V(q) = \frac{\vec{q}^{2}}{\vec{q}^{2}+m^{2}_{\pi}}.
\end{equation}
In UCOM, a unitary correlation operator is introduced as a shift operator, $C$, where $\psi = C\phi$ represents 
correlated states and $C^{\dagger} H C \phi =  E\phi$ represents the equation of the uncorrelated states, $\phi$.
This unitary correlation operator moves two particles away from each other whenever they are within a short distance 
due to the repulsive core \cite{feldmeier}.  We apply this unitary correlation operator for only the central part of 
the pion-exchange interaction within the approximation of taking terms up to two-particle correlated operators.
The interaction with form factor in coordinate representation is
\begin{eqnarray}
V(r) &=& \frac{1}{(2\pi)^{3}} \int d\vec{q} V(q) 
\Big{(} \frac{\Lambda^{2} - m^{2}_{\pi}}{\Lambda^{2}+\vec{q}^{2}} \Big{)}^{2} \exp(i\vec{q}\cdot\vec{r}) \\ \nonumber
&=& - \frac{1}{4\pi} 
\Big{(} m^{2}_{\pi}\frac{\exp(-m_{\pi}r)}{r}
- \frac{\Lambda^{2}_{2} - m^{2}_{\pi}}{\Lambda^{2}_{2}-\Lambda^{2}_{1}} 
\Lambda^{2}_{1} \frac{\exp(-\Lambda_{1}r)}{r} 
+ \frac{\Lambda^{2}_{1} - m^{2}_{\pi}}{\Lambda^{2}_{2}-\Lambda^{2}_{1}} 
\Lambda^{2}_{2}\frac{\exp(-\Lambda_{2}r)}{r} \Big{)},
\end{eqnarray}
where $\Lambda_{1} = \Lambda + \epsilon$, 
$\Lambda_{2} = \Lambda - \epsilon$, and $\epsilon$ is taken to be 10 MeV to express differential operation with respect to 
$\Lambda$ \cite{machleidt}. The correlated operator is given as
\begin{equation}
\widetilde{V}(r) = C^{\dagger} V(r) C = V(R_{+}(r)),
\end{equation}
where $R_{+}(r) = r + \alpha (\frac{r}{\beta})^{\eta} \exp(-\exp(\frac{r}{\beta}))$.  
The parameters, $\alpha$ = 0.94 fm, $\beta$ = 1 fm, and $\eta$ = 0.37 are used in Ref. \citen{feldmeier}.  
The correlated operator in momentum representation, $\widetilde{V}(q)$, is obtained by Fourier transformation 
of $\widetilde{V}(r)$.

\begin{figure}[t]
\includegraphics[width=6.9cm,clip]{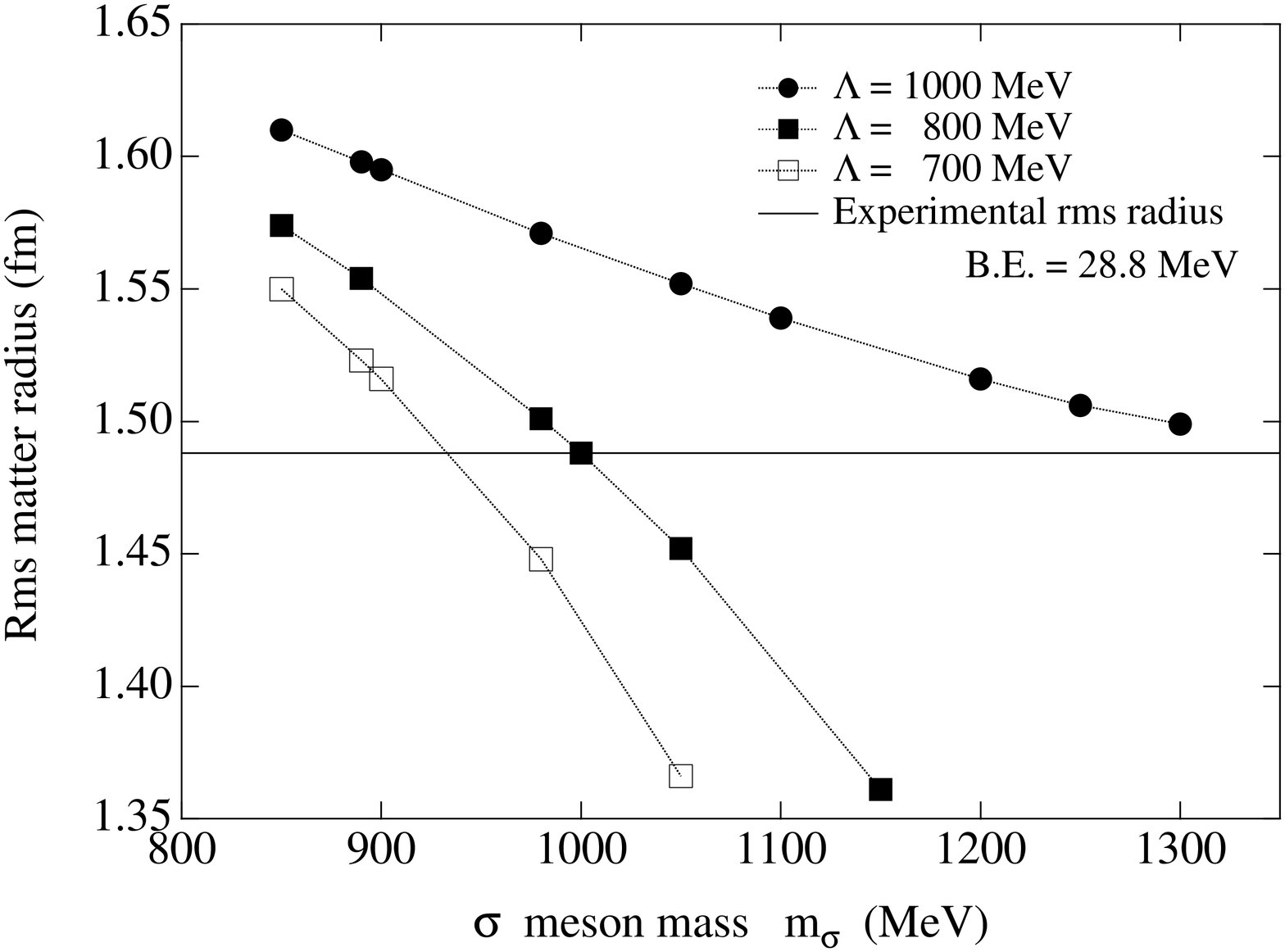}
\includegraphics[width=6.9cm,clip]{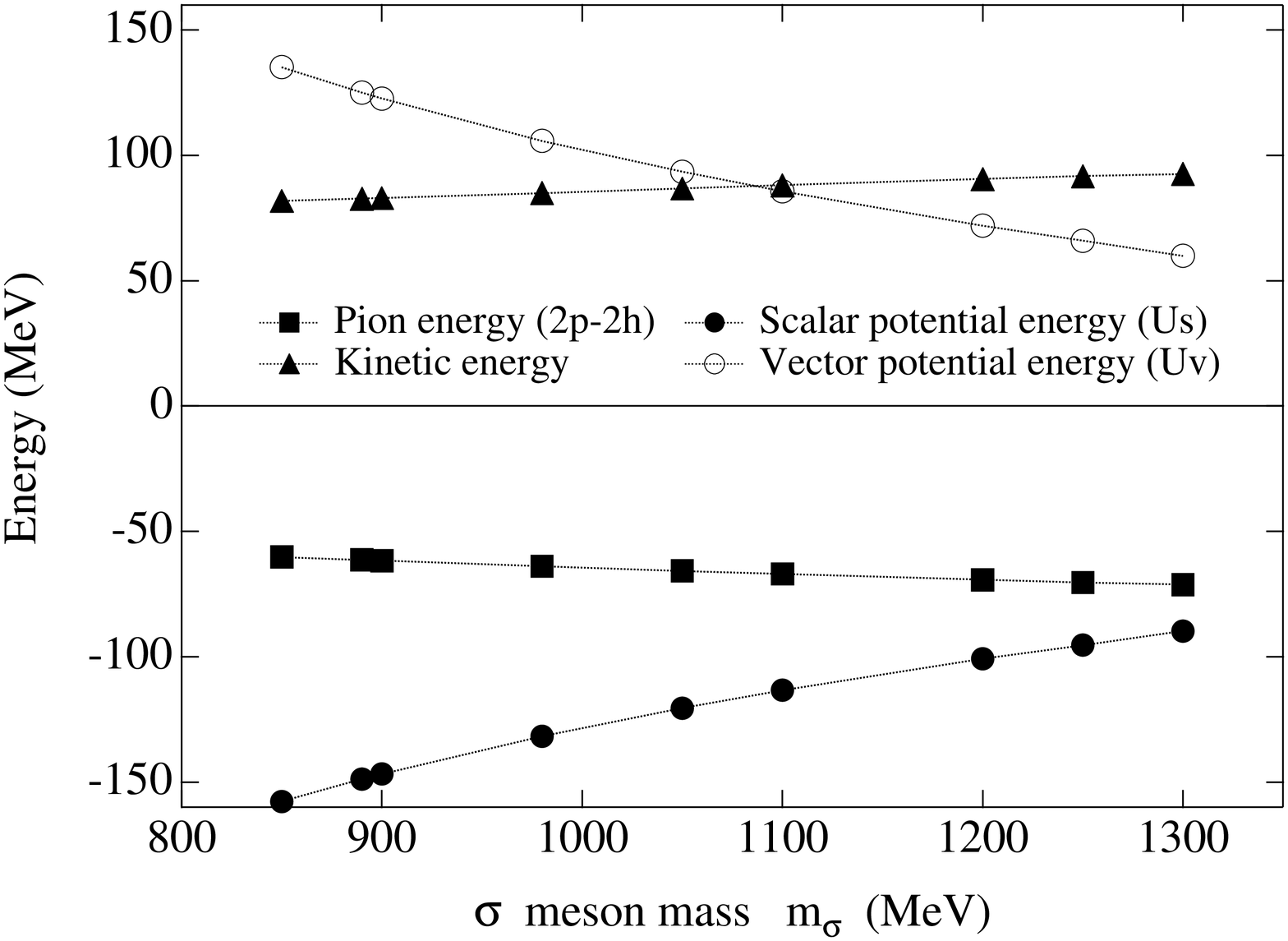}
\caption{\label{fig9} Root-mean-square matter radius of $^4$He for several cases of cutoff momentum, 
$\Lambda$ as function of the $\sigma$ meson mass in the left panel. The solid circle represents the result for 
$\Lambda$ = 1000 MeV, the solid square for $\Lambda$ = 800 MeV, and the open square for $\Lambda$ = 700 MeV.  
The right-hand panel shows energy components as functions of the $\sigma$ meson mass for $\Lambda$ = 1000 MeV.
The pionic quantum number is taken up to $J^{\pi}_{\rm max}$ = 6$^{-}$.   The number of Gaussians is 8, 
where $b_i$ = 0.4, 0.6, 0.8, 0.9, 1.0, 1.1, 1.4, 2.0 fm.}
\end{figure}
We show the mean square radius of $^4$He in Fig. \ref{fig9}.  To reproduce both the binding energy and 
the rms matter radius simultaneously, we need a significantly large $\sigma$ meson mass as compared with 
the case of the phenomenological parameter set, for example, the Walecka model \cite{walecka,sugahara}.  
As the $\sigma$ meson mass becomes heavier, the attraction due to the $\sigma$ field decreases and 
the energy contribution from the pion-exchange interaction increases.  The $\sigma$ meson mass should be 
$m_{\sigma}$ = 1300, 1000, and 940 MeV in cases with $\Lambda$ = 1000, 800, and 700 MeV, respectively, 
to reproduce both the rms matter radius and binding energy.   In the right-hand panel of Fig. \ref{fig9}, we show 
the energy components for $^{4}$He in the case of $\Lambda$ = 1000 MeV.  Around 70$\%$ of the total 
attractive potential comes from the pion-exchange interaction when the binding energy and rms matter radius 
are reproduced simultaneously.  This amount of energy contribution due to the pion-exchange interaction is 
in good agreement with the result of the VMC method for the light nuclei obtained by the Argonne-Illinois 
group \cite{pandharipande}.  The energy contribution due to the pion-exchange interaction decreases 
as the cut-off momentum becomes smaller.

\begin{figure}[t]
\centering
\includegraphics[width=6.9cm,clip]{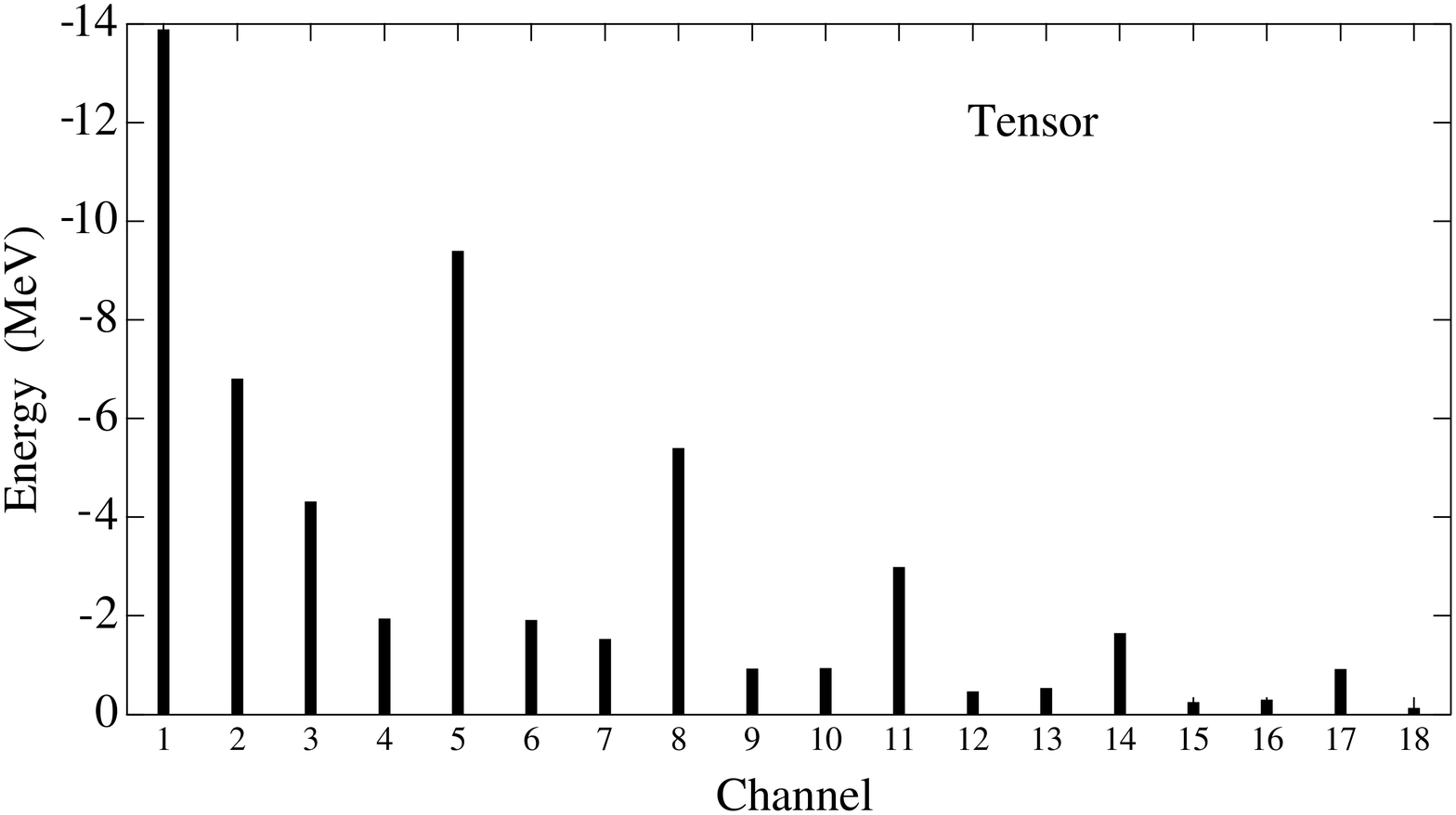}
\includegraphics[width=6.9cm,clip]{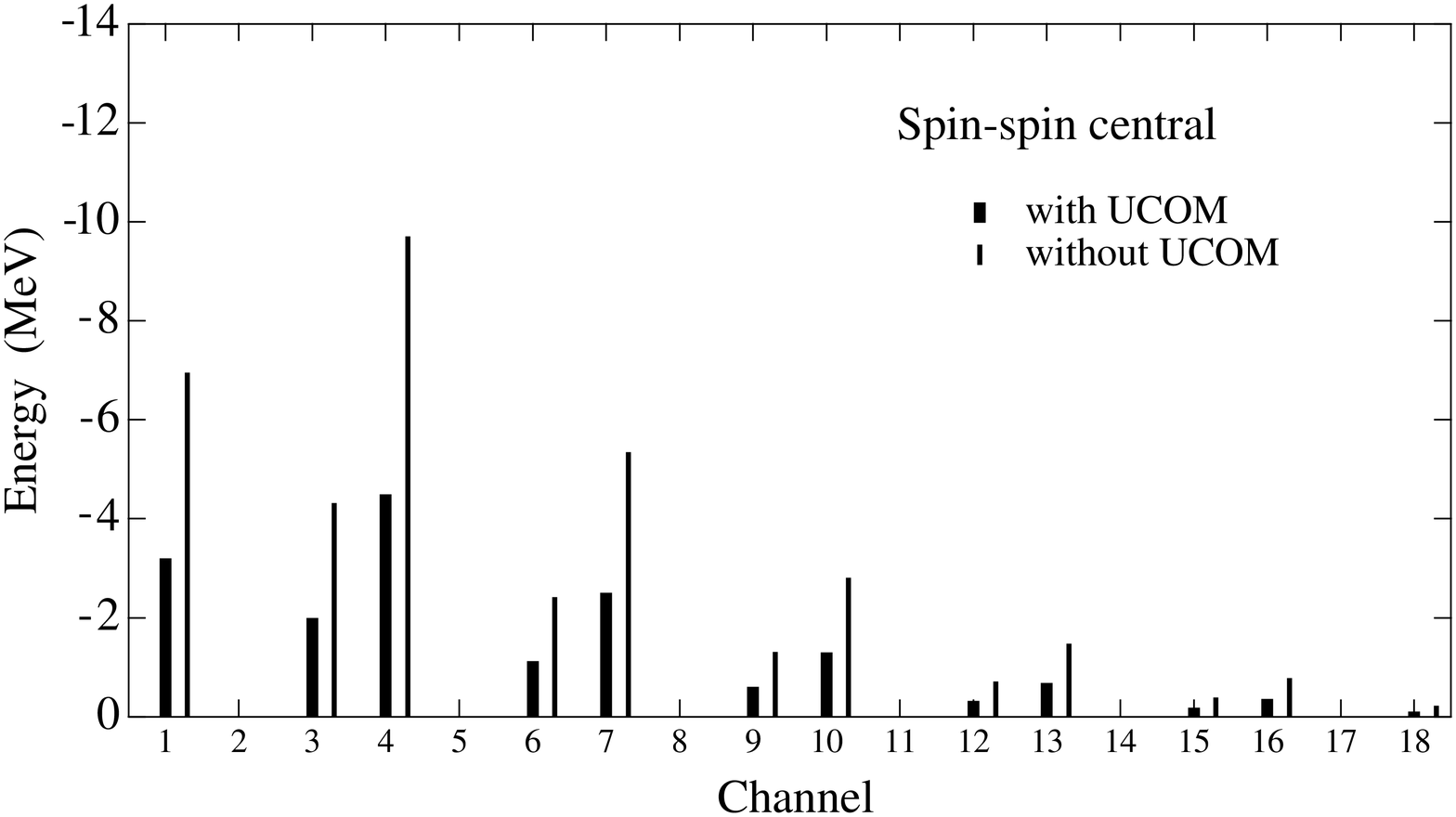}
\caption{\label{fig10} 
Energy contributions of the tensor part and spin-spin central part of the pion-exchange interaction 
for various 2p-2h states with $\Lambda$ = 1000 MeV.  The left-hand panel shows the energy from 
the tensor part. In the right-hand figure is shown the energy from the spin-spin central part.  The thin bars 
in the right-hand panel represent the energy without the UCOM operation.  The binding energy and rms matter 
radius are reproduced simultaneously by adjusting the $\sigma$ meson mass and $\omega$-nucleon 
coupling constant.  Each number shown in the horizontal axis represents the 2p-2h channels, where 
1:$(p_{1/2})^{2}$, 2:$(1s_{1/2})(d_{3/2})$, 3:$(d_{3/2})^{2}$, 4:$(p_{3/2})^{2}$, 5:$(p_{3/2})(f_{5/2})$,
6:$(f_{5/2})^{2}$, 7:$(d_{5/2})^{2}$, 8:$(d_{5/2})(g_{7/2})$, 9:$(g_{7/2})^{2}$, 10:$(f_{7/2})^{2}$,
11:$(f_{7/2})(h_{9/2})$, 12:$(h_{9/2})^{2}$, 13:$(g_{9/2})^{2}$, 14:$(g_{9/2})(i_{11/2})$, 
15:$(i_{11/2})^{2}$, 16:$(h_{11/2})^{2}$, 17:$(h_{11/2})(j_{13/2})$, 18:$(j_{13/2})^{2}$.
The number of Gaussians is 8, where $b_i$ = 0.4, 0.6, 0.8, 0.9, 1.0, 1.1, 1.4, 2.0 fm.}
\end{figure}
We calculate the energy contributions from both the tensor part and spin-spin central part of the pion-exchange interaction 
given in Eq. (3.2) and show the results in Fig. \ref{fig10}.  When we take into account the effect of the short-range repulsion 
by using the UCOM prescription for the spin-spin central part, the ratio of the energy from the spin-spin central part to the one 
from the tensor part is around 1/3.  Around 50$\%$ of the total attractive potential comes from the tensor part, 
and about 20$\%$ comes from the spin-spin central part.  We compare these energy components with the result obtained
by the Argonne-Illinois group \cite{pandharipande}.  We obtain good agreement with their result.  In particular, we can reduce 
the energy of the spin-spin central part of the $(p_{3/2})^{2}$ state.

\subsection{Density distribution}
\begin{figure}[b]
\centering
\includegraphics[width=6.9cm,clip]{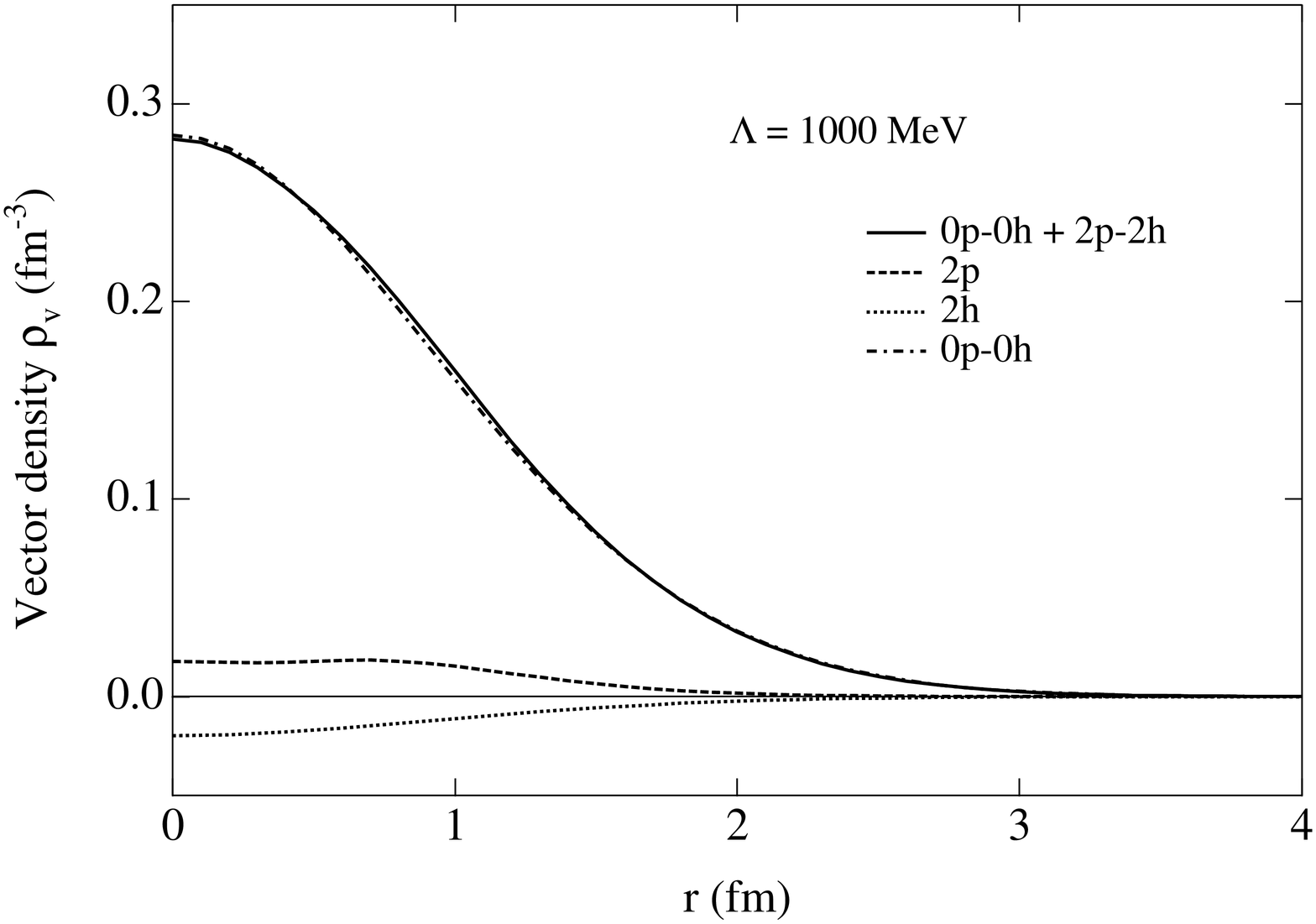}
\includegraphics[width=6.9cm,clip]{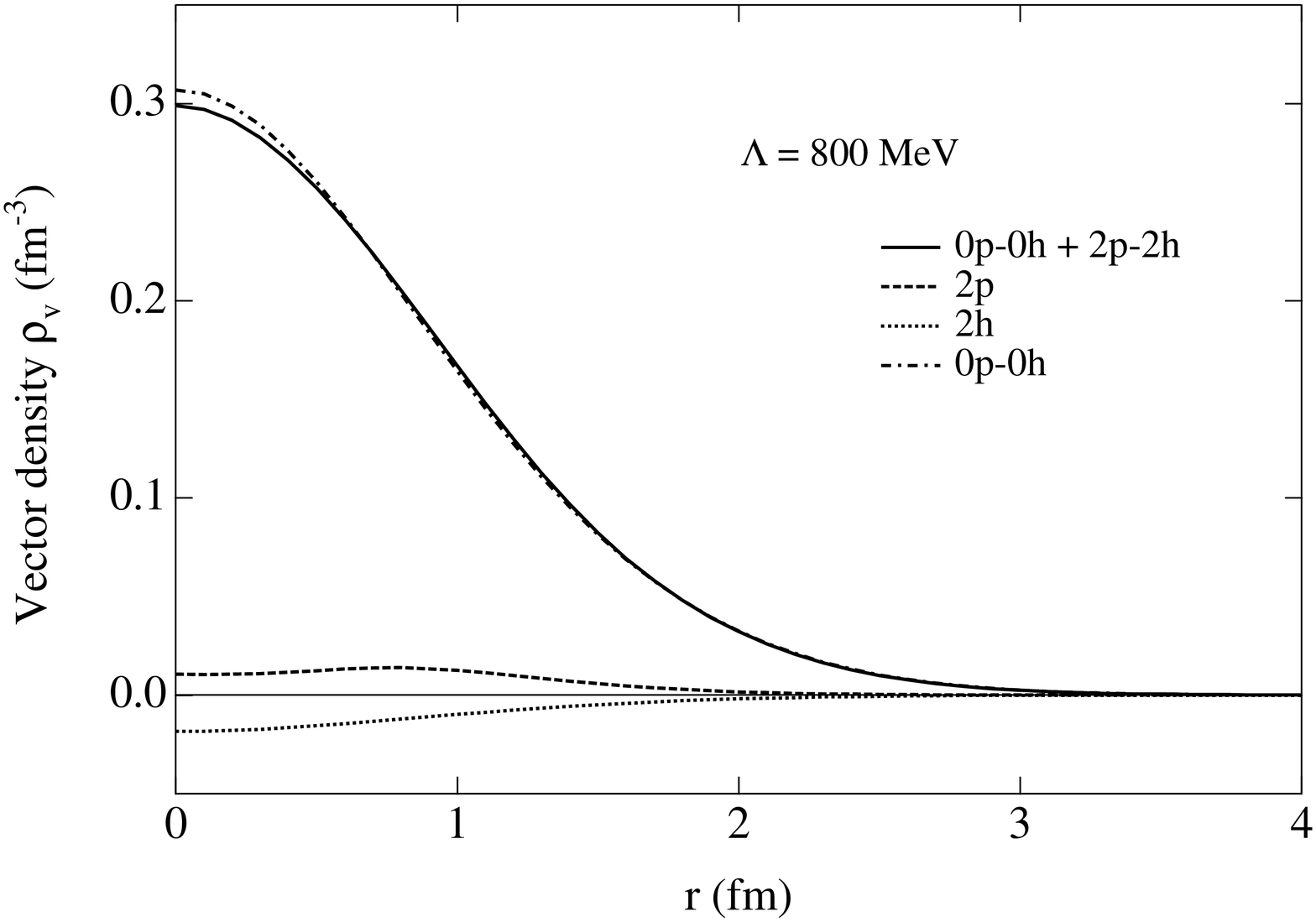}
\caption{\label{fig11} The vector density distribution is shown as a function of the radial distance, $r$, 
in the left panel for $\Lambda$ = 1000 MeV and in the right panel for $\Lambda$ = 800 MeV.  The dash-dotted 
curve represents the density of the 0p-0h RMF ground state, while the dashed curve represents the density 
distribution of the two-particle state and the dotted curve represents the density distribution of the two-hole state.  
The solid curve represents the density distribution of total ground state wave function, 0p-0h + 2p-2h.   
For both cases, the parameters are adjusted to reproduce the binding energy and rms matter radius simultaneously.  
The pionic quantum number is taken up to $J^{\pi}_{\rm max}$ = 6$^{-}$.   The number of Gaussians is 8, 
where $b_i$ = 0.4, 0.6, 0.8, 0.9, 1.0, 1.1, 1.4, 2.0 fm.}
\end{figure}
In Fig. \ref{fig11}, we show the vector density distribution for $^{4}$He for $\Lambda$ = 1000 MeV 
in the left panel and for 800 MeV in the right panel.  In both cases, the binding energy and rms matter radius 
are reproduced simultaneously.  The dashed and dotted curves represent the densities of two-particle component 
and two-hole component of 2p-2h states, respectively.  The integrated value of the density of the two-particle 
component is twice the probability of the 2p-2h states, $\sum_{i} \alpha^{\ast} _{i} \alpha _{i}$.  The amount 
of this probability is around 14$\%$ for $\Lambda=1000$ MeV.  The amount of twice of this probability 
of the density distribution moves from the 0p-0h RMF ground state to the two-particle excited state through 
the pionic correlation due to the pion-exchange interaction.  This 2p-2h contribution changes the density 
distribution of $^4$He.  The central part of the density distribution is reduced and the outside region 
at $r > 2$ fm of the density distribution is also reduced, and consequently, the distribution becomes spatially 
compact.  A peak of the fluctuation density distribution of the particle-state arises at around 0.8 fm, 
and this fact indicates the important property of the pion-exchange interaction.  The density distribution 
of two-particle states is spatially compact reflecting the pseudo-scalar nature of the pion-exchange interaction.

\begin{figure}[b]
\centering
\centerline{\includegraphics[width=8.0cm,clip]{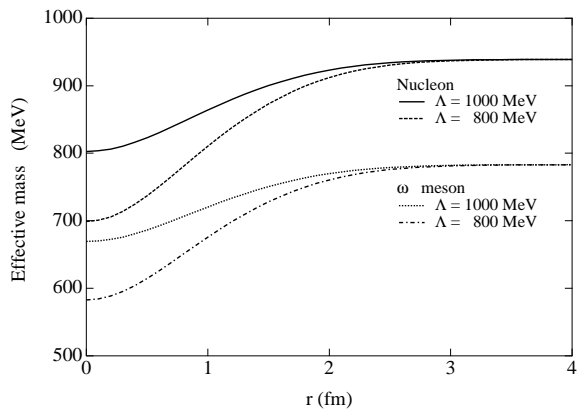}}
\caption{\label{fig12} 
Effective masses of the nucleon and $\omega$-meson for $\Lambda$ = 1000 and 800 MeV.  For both cases, 
the parameters are adjusted to reproduce the binding energy and rms matter radius simultaneously.  
The pionic quantum number is taken up to $J^{\pi}_{\rm max}$ = 6$^{-}$.  The number of Gaussians is 8, 
where $b_i$ = 0.4, 0.6, 0.8, 0.9, 1.0, 1.1, 1.4, 2.0 fm.}
\end{figure}
We show in Fig. \ref{fig12} the effective masses for the nucleon and $\omega$ meson as a function of the nuclear 
radius for $\Lambda$ = 1000 and 800 MeV.  The effective masses are given for the nucleon and $\omega$ meson as
\begin{eqnarray}
M^{\ast} &=& M + g_{\sigma}\sigma, \\ \nonumber
{m^{\ast}}_{\omega} &=&  m_{\omega} + \widetilde{g}_{\omega} \sigma.
\end{eqnarray}
The pionic energy contribution becomes important in cases of large cutoff momentum.  For large $\Lambda$, 
the $\sigma$ meson mass should be larger and the contribution of the $\sigma$ meson becomes small.  
Hence, the nucleon and $\omega$ meson masses  are larger for larger $\Lambda$.  

\section{Summary}
We have formulated the RCMF model for finite nuclei based on the relativistic single-particle framework.  
This RCMF model is a natural extension of the CPPRMF model to include higher pionic quantum numbers 
of particle-hole states, $J^{\pi} = 1^{+}, 2^{-}, 3^{+}$, .... , to take into account the full strength of 
the pion-exchange interaction.  The CPPRMF model corresponds to the $J^{\pi} = 0^{-}$ case in the 2p-2h states 
in the RCMF model, which reflects the surface contribution of the pion-exchange interaction.   The inclusion of 
higher multipoles comes from the necessity to include the volume contribution of the pion-exchange interaction.  
We have applied the RCMF model to the $^{4}$He nucleus as a pilot calculation and studied the energy convergence 
and role of pion in the nuclear structure.  The RCMF model has three free parameters in the sigma model Lagrangian. 
They are the $\sigma$ meson mass, $m_{\sigma}$, the $\omega$-nucleon coupling constant, $g_{\omega}$, 
and the cutoff momentum, $\Lambda$.  These parameters are fixed so as to reproduce the binding energy 
and radius of $^4$He and the contribution of the tensor interaction of the pion-exchange interaction.  We have 
obtained the energy convergence when we take into account the pionic quantum number up to 
$J^{\pi}_{\rm max}$ = 6$^{-}$.

The Gaussian ranges are taken as the energy variational parameters.  The spatially compact distribution 
of two-particle states is obtained explicitly at the point where the minimum of the total energy is realized.  
This fact reflects the involvement of the high-momentum component in the wave function.  This framework 
can describe the pseudo-scalar nature of the pion-exchange interaction in this variational method.  
When we increase the number of Gaussians to expand particle states, the pion energy gain increases 
by around 20$\%$ from the case of one-range Gaussian.  In this framework, the number of Gaussian 
expansions does not need to be high, because the Gaussian range, $b_{i}$, is taken as the variational parameter.
We can take into account the full strength of the pionic correlation in the intermediate range by using this framework.

To reproduce both binding energy and rms matter radius simultaneously, we need a significantly large 
$\sigma$ meson mass as compared with that of the phenomenological parameter set, for example, 
the Walecka model \cite{walecka, sugahara}.  As the $\sigma$ meson mass becomes heavier, the attraction 
due to the $\sigma$ field decreases and the energy contribution from the pion-exchange interaction increases.  
The $\sigma$ meson mass is taken to be $m_{\sigma}$ = 1300, 1000, and 940 MeV in the case with 
$\Lambda$ = 1000, 800, and 700 MeV, respectively. 

Around 70$\%$ of the total attractive potential comes from the pion-exchange interaction for 
$\Lambda$ = 1000 MeV, and at this moment, the binding energy and rms matter radius are reproduced 
simultaneously.  This amount of energy contribution due to the pion-exchange interaction is in good agreement 
with the result of the VMC method for the light nuclei obtained by the Argonne-Illinois group \cite{pandharipande}.   
We have calculated various energy contributions from both spin-spin central part and tensor part of the 
pion-exchange interaction given in Eq. (3.2). Both parts give almost the same amount of energy contribution 
when we do not take care of the short-range repulsion.  When we take into account the effect of the short-range 
repulsion by using the UCOM prescription for the spin-spin central part, the ratio of the energy contribution 
from the spin-spin central part and tensor part is around 1/3.  Around 50$\%$ of the total attractive potential 
comes from the tensor part, and about 20$\%$ comes from the spin-spin central part. We compare these energy 
constituents with the result obtained by Argonne-Illinois group \cite{pandharipande}.  We obtain good 
agreement with their result.

As for the energy contribution due to the tensor part, the most important two-particle state in the 2p-2h channels 
is the $(p_{1/2})^{2}$.  The next ones are the $(p_{3/2})(f_{5/2})$, $(1s_{1/2})(d_{3/2})$, and $(d_{3/2})^{2}$ 
in order of importance.  As the energy contribution due to the spin-spin central part, the $(p_{3/2})^{2}$ state 
leads to the most important contribution. The next ones are the $(p_{1/2})^{2}$, $(d_{5/2})^{2}$, $(d_{3/2})^{2}$ 
in order of importance.

In this work, the lower radial wave function is connected with the upper radial wave function with the plane wave 
relation through $\frac{1}{2M}$, to minimize the number of variational parameters. We would like to remove 
this constraint for small components as variational parameters.  Furthermore, we have to include the 
exchange (Fock)-term in this framework, since we describe the problem of the many-body system.  
These points will be improved in forthcoming works.  We shall proceed to calculate heavier nuclei to ensure 
the validity of our statement in the previous work, where the pion plays a role in the formation of the 
$jj$-magic shell structure \cite{ogawa3}.  We have shown the effective masses of the nucleon and 
$\omega$ meson inside the nucleus.  We would like to consider the relation between pion and $\sigma$-field 
for the purpose of the discussion of the chiral symmetry in finite nucleus in future work.  We have applied 
UCOM prescription to treat properly the short-range repulsion for the spin-spin central part of the pion-exchange
interaction.  We have to develop a dynamical framework to determine the amount of short-range correlation 
by explicitly including short-range repulsive interaction.


\section*{Acknowledgements}
We are grateful to Dr. T. Myo and Prof. K. Ikeda for useful discussion on the tensor-optimized shell model.  
Y. O. is thankful to the members of the RCNP theory group for fruitful discussions.  This work is supported 
in part by Grant-in-Aid for Scientific Research (C) No.18540269 from the Ministry of Education, Culture, 
Sports, Science and Technology of Japan.

%

\end{document}